\documentclass[lettersize,journal]{IEEEtran}
\usepackage{amsmath,amssymb,amsfonts}
\usepackage{algorithmic}
\usepackage{algorithm}
\usepackage{array}
\usepackage[caption=false,font=normalsize,labelfont=sf,textfont=sf]{subfig}
\usepackage{textcomp}
\usepackage{stfloats}
\usepackage{url}
\usepackage{verbatim}
\usepackage{graphicx}
\usepackage{xcolor}
\usepackage{hyperref}

\usepackage{amsthm}
\usepackage{cite}
\usepackage{array, makecell}
\usepackage{mathtools}
\newtheorem{theorem}{Theorem}

\newcommand\halfwidth{1.0} 

\usepackage[font=small,labelfont=bf]{caption}
\begin{document}

\title{Robust Eavesdropping in the Presence of Adversarial Communications for RF Fingerprinting}

\author{Andrew Yuan and Rajeev Sahay

\thanks{A. Yuan and R. Sahay are with the Department of Electrical and Computer Engineering, University of California San Diego, San Diego, CA, 92093 USA. E-mail: \{a1yuan,r2sahay\}@ucsd.edu.}
\thanks{This work was supported in part by the University of California San Diego Academic Senate under grant RG114404 and in part by the National Science Foundation (NSF) under grant ECCS-2512912.}}



\maketitle

\begin{abstract} 

Deep learning is an effective approach for performing radio frequency (RF) fingerprinting, which aims to identify the transmitter corresponding to received RF signals. However, beyond the intended receiver, malicious eavesdroppers can also intercept signals and attempt to fingerprint transmitters communicating over a wireless channel. Recent studies suggest that transmitters can counter such threats by embedding deep learning-based transferable adversarial attacks in their signals before transmission. In this work, we develop a time-frequency-based eavesdropper architecture that is capable of withstanding such transferable adversarial perturbations and thus able to perform effective RF fingerprinting. We theoretically demonstrate that adversarial perturbations injected by a transmitter are confined to specific time-frequency regions that are insignificant during inference, directly increasing fingerprinting accuracy on perturbed signals intercepted by the eavesdropper. Empirical evaluations on a real-world dataset validate our theoretical findings, showing that deep learning-based RF fingerprinting eavesdroppers can achieve classification performance comparable to the intended receiver, despite efforts made by the transmitter to deceive the eavesdropper. Our framework reveals that relying on transferable adversarial attacks may not be sufficient to prevent eavesdroppers from successfully fingerprinting transmissions in next-generation deep learning-based communications systems.

\end{abstract}

\begin{IEEEkeywords}
Adversarial attacks, deep learning, eavesdropping, physical-layer communications, rf fingerprinting
\end{IEEEkeywords}

\section{Introduction}

\IEEEPARstart{A}{s} the number of devices requiring wireless resources has increased, the Internet of Things (IoT) has become the backbone for supporting the vast flux of wireless traffic. The widespread usage of IoT devices has led to the development of security measures to protect wireless networks from adversarial actors. In particular, these measures include, among other protocols, authenticating devices in wireless networks prior to servicing them. For example, cryptographic approaches have been deployed, distributing unique keys at each node to prevent non-authenticated sources from intercepting wireless transmissions. Yet, such software-based security frameworks are susceptible to attacks such as denial of service (DoS) \cite{dos}, MAC/IP address spoofing \cite{macattack}, and physical tampering \cite{phys_tampering}.

To address the vulnerabilities of software-based security frameworks, hardware-based approaches have recently emerged to perform device authentication. In this capacity, \emph{radio frequency (RF) fingerprinting (RFF)} has attracted attention for its promising ability to identify and authenticate devices from received signals. Specifically, the inherent imperfections of hardware components are captured in the channel state information (CSI) of received wireless signals and serve as unique identifiers, which can effectively \emph{fingerprint} (i.e., identify) their corresponding transmitter. However, traditional approaches for RFF, such as maximum likelihood estimation (MLE) classifiers \cite{MLE1,MLE2,MLE3} that operate on higher order statistical moments as features, require \emph{a priori} knowledge of the channel and communication protocols, and they often do not have the capacity to generalize to unfamiliar conditions. Alternatively, deep learning has emerged as a promising paradigm for RFF due to its ability to efficiently distinguish RF fingerprints directly from raw in-phase and quadrature (IQ) time samples without requiring the calculation of hand-crafted features in dynamic and congested wireless networks \cite{DLRF8, DLRF9}.

Standard RFF wireless communications environments consist of a transmitter that broadcasts a wireless signal and a receiver that employs a deep learning model to fingerprint the received signal using its complex baseband representation \cite{DLRF1,DLRF2}. In addition, such wireless mediums can also consist of adversaries aiming to compromise the end-to-end RFF process. Although transmitters' fingerprints cannot be mimicked, due to the challenge of replicating the hardware defects of another device, it is common for rogue receivers to eavesdrop on sensitive information. Specifically, due to the broadcast nature of wireless communications, any adversary can effectively sense and locally fingerprint a transmitted signal, reducing the covertness in a communication channel. In a typical over-the-air wiretapping scheme, an eavesdropper may choose to be either active or passive. Active eavesdroppers intercept and modify the message within the transmission, but expose their existence \cite{eav1}. In contrast, passive eavesdroppers merely receive and decode the intercepted transmission, which may be degraded by environmental factors (e.g., path loss, shadowing, multi-path fading, etc.) \cite{passiveeavsdropper}. 

Through passively capturing RF signatures, malicious eavesdroppers may compile a directory of authorized and prohibited agents within a secured entity, giving knowledge of potential associations between agents which may expose the structure, behavior, and other sensitive information regarding the private network of trusted devices. With the double-edged sword of using RFF to track a user’s device through their received signal strength (RSS), the additional information of the trusted device's location allows for a more carefully coordinated attack by the malicious eavesdropper. For instance, \cite{rfftracking1} demonstrated how passive RFF collection allows for tracking of WiFi devices in commercial and military environments, compromising physical security breaches and user privacy. In a similar motivation, \cite{rfftracking2} through an attacker’s perspective, used RFF to infer the temporal and geographical status of vehicles, drones, and operation movement. To address the dire consequences of passive eavesdropping, this paper assumes that the inactive eavesdropper does not manipulate the transmission but instead attempts to perform transmitter authentication on a database of trusted devices within a private wireless network using their intercepted RF signatures.

Prior work has shown that transmitters can effectively evade eavesdroppers by injecting adversarial deep learning perturbations into their transmitted signals \cite{amc_eaves,amc_eaves2,amc_eaves3,amc_eaves4,evs1,evs3,evs4,evs5}. Due to the transferability property of adversarial deep learning attacks \cite{transfer1,transfer2,transfer3}, adversarial attacks injected by the transmitter degrade the performance of deep learning-based RFF classifiers at an eavesdropper even when the eavesdropper has imperfect knowledge about the receiver's classification architecture. To mitigate performance degradation from adversarial perturbations at the receiver, the receiver in collaboration with the transmitter, filters the adversarial perturbation from the received signal before performing RFF, a method known to be effective in environments where the receiver is aware of the transmitter’s perturbation approach \cite{DAE2,MIMODAE}. 

In this work, we focus on the perspective of the eavesdropper in a deep learning-based RFF framework. Specifically, we develop an approach for an eavesdropper to effectively
perform RFF in a wireless environment in which the transmitters and a receiver are using adversarial attacks to deceive the eavesdropper. We consider a private network consisting of nominal trusted and untrusted transmitting nodes broadcasting a potentially corrupted message to a receiver, which reconstructs the original fingerprint from the perturbations, and uses the in-phase and quadrature (IQ) samples to determine whether the device is trustworthy or untrustworthy. Simultaneously, a malicious eavesdropper, aware of potential signal perturbation fingerprints crafted on a receiver classifier, intercepts the transmitted signal, and mitigates the influence of the perturbation (using the benchmark schemes as detailed in Sec. \ref{exp_setup}). By determining when and where the secured entity (the receiver) is servicing an authorized device, the eavesdropper may gain sensitive information of the private network's behavior, allowing for malevolent intentions such as location tracking. We considered two scenarios: the eavesdropper has limited knowledge (Environment A) and no knowledge (Environment B) of the classification architecture at the receiver, which affects the transferability of the adversarial attack on the eavesdropper model. To this end, we develop a time-frequency-based approach, which we theoretically and empirically show to minimize the effect of additive perturbations at the eavesdropper, resulting in RFF performance similar to the receiver at the eavesdropper. Our approach demonstrates that deep learning-based eavesdroppers are difficult to evade through the injection of transferable adversarial attacks when the transmitter is unaware of the eavesdropper's time--frequency defense strategy thus revealing a major deep learning-based RFF vulnerability, which is particularly concerning in security sensitive applications such as military environments.

\textbf{Outline and Summary of Contributions:} Compared to related work in RFF under adversarial conditions (discussed in Sec. \ref{relaeted_wrks}), we make the following contributions (in the stated sections):
\begin{enumerate}
    \item \textbf{A robust RFF eavesdropper architecture} (Sec. \ref{sig_mod} -- \ref{sec:ee}): To the best of our knowledge, we are the first to develop a novel methodology for an eavesdropper to perform accurate RFF in the presence of intentional adversarial perturbations injected by the transmitter.
    
    \item \textbf{Theoretical analysis of eavesdropping framework} (Sec. \ref{thm}):  We theoretically demonstrate the feasibility of our proposed RFF eavesdropping architecture using temporal-spatial analysis, which effectively eliminates effects of adversarial perturbations that are specifically generated to deceive an eavesdropper.
    
    \item \textbf{Empirical evaluation on real-world data} (Sec. \ref{exp_setup} -- Sec. \ref{pgd_res}): Using a real world RFF dataset, we show that, using our approach, the eavesdropper is able to recover and identify the trustworthiness of devices at a higher performance compared to existing benchmark approaches for adversarial attack mitigation on multiple perturbation designs across a range of potencies in various signal environments.

\end{enumerate}




\section{Related Works} \label{relaeted_wrks}

Deep learning, using both convolutional neural network (CNN) and recurrent neural network (RNN) architectures, has been overwhelmingly shown to perform effective RFF on IQ samples \cite{DLRF1,DLRF2,DLRF3,DLRF4} as well as on constellation diagrams formed from IQ samples \cite{DLRF5}. A common pitfall of deep learning classifiers, however, is their vulnerability to ``blindspots" located in high dimensional spaces which drastically sabotage their performances. In domains such as computer vision \cite{cvadv_attack} and speech processing \cite{speech_advattack}, imperceptible adversarial attacks, are crafted to exploit the inherent vulnerabilities of deep learning models. In device recognition, \cite{rffidattack1,rffidattack2,rffidattack3} have demonstrated the detrimental effects of adversarial attacks in practical scenarios where an attacker crafts a perturbation despite having limited knowledge of the classifier and imperfect CSI. In response, several defenses have been investigated to mitigate the effect of adversarial attacks in RFF. \cite{rffiddefense1,rffiddefense2, rffiddefense3} increased the classifier robustness through various forms of adversarial training, while sacrificing accuracy on clean IQ samples. In addition, other defenses have also been proposed for wireless communications tasks outside of RFF. For example, \cite{amcdefense1, amcdefense2} implemented autoencoders capable of denoising adversarial perturbations injected into modulated signals as a means to fool Automatic Modulation Classification (AMC) models. Our work can be considered orthogonal to such approaches in that we are not interested in developing a defense to improve the robustness of RFF models but rather we show that adversarial attacks crafted specifically to deceive eavesdroppers in a wireless channel can be circumvented.


The notion of generating adversarial perturbations to fool unwanted eavesdroppers equipped with RFF classifiers have been investigated. In \cite{stayconnected}, a random phase is embedded to each subcarrier in a WiFi orthogonal frequency-division multiplexing (OFDM) system where non-linear phase errors are used for RFF identification, demonstrating robustness against statistically based attacks. \cite{hist} develops a novel adversarial defense by transforming the perturbed RF signatures into constellation images which improves the reliability against adversarial attacks at the cost of a lower baseline performance. A similar motivation appears in \cite{rffeavesdropper}, where from a transmitter perspective, an artificially generated perturbation successfully misleads an eavesdropper by modifying the pilot signal prior to transmitting. Of the aforementioned efforts, our work is most closely related to \cite{rffeavesdropper}. However, we consider an eavesdropping RFF scheme robust to the transmitter’s RF concealment designed to authenticate, rather than identify, known and unknown transmitters that enter and leave a wireless network in an ad-hoc environment. 

\section{Methodology}

\begin{figure}[t] 
	\centering
	\includegraphics[width=\halfwidth\columnwidth]{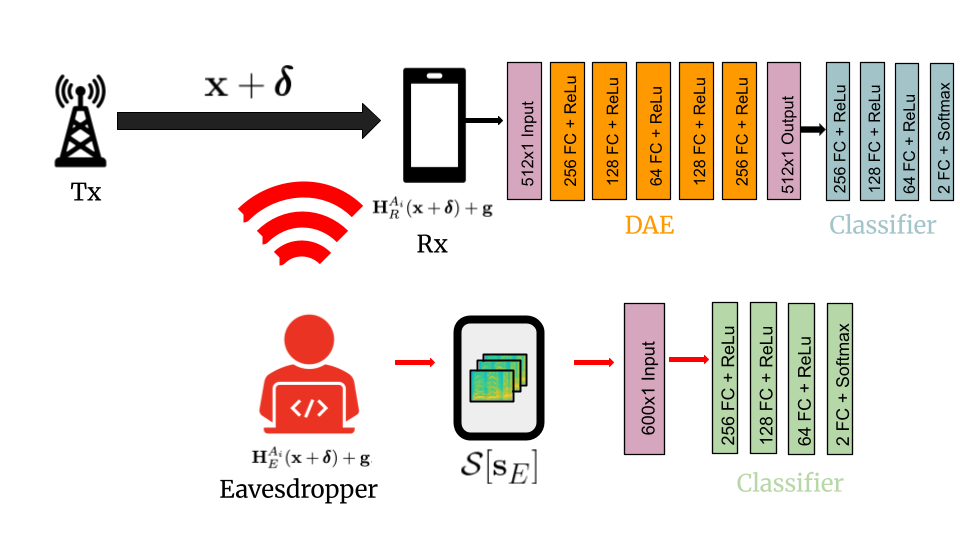}
	\caption{Our considered system diagram consisting of an RF Fingerprinting architecture in the presence of an eavesdropper.}
	\label{sys_diag}
\end{figure}

In this section, we model our methodology for performing effective deep learning-based device fingerprinting from an eavesdropper on an intercepted signal containing adversarial perturbations. We begin by defining our signal, channel, and classifier models used for RF fingerprinting by the transmitter, receiver, and eavesdropper (Sec. \ref{sig_mod} and \ref{clf_mod}). Next, we define the adversarial environment created between the transmitter and receiver to fool an eavesdropper (Sec. \ref{sec:ee}). Finally, we develop our method for deep learning-based eavesdroppers to withstand such adversarial environments and show the theoretical performance guarantees of our framework (Sec. \ref{thm}). Our overall framework is shown in Fig. \ref{sys_diag}.


\subsection{Signal and Channel Modeling} \label{sig_mod}
As shown in Fig. \ref{sys_diag}, we consider a wireless communication environment consisting of a transmitter, receiver, and eavesdropper. Both the receiver and the eavesdropper aim to perform RF fingerprinting on their received signal and classify it as either a trusted or untrusted device. The objective of the transmitter is to transmit a signal that prevents accurate fingerprinting at the eavesdropper while maintaining accurate fingerprinting at the receiver. Simultaneously, the objective of the eavesdropper is to perform accurate fingerprinting on its intercepted signal despite the measures taken by the transmitter to deceive the eavesdropper.


In our framework, we consider a set of $\mathcal{A} = \{A_{1}, A_{2}, \ldots, A_{|\mathcal{A}|}\}$ transmitters. We further decompose $\mathcal{A}$ into $\mathcal{A} = \mathcal{T} \hspace{1mm} \cup \hspace{1mm} \mathcal{U}$, where $\mathcal{T}$ is the set of trusted transmitters, $\mathcal{U}$ is the set of untrusted transmitters, and $\mathcal{T} \hspace{1mm} \cap \hspace{1mm} \mathcal{U} = \emptyset$. Each transmitter wirelessly transmits $\mathbf{x} = [x[0], \cdots, x[\ell-1]]$ through a channel $\mathbf{h}_{R}^{A_{i}} \in \mathbb{C}^{\ell}$, where $\mathbf{h}_{R}^{A_{i}} \in [h_{R}^{A_{i}}[0], \cdots, h_{R}^{A_{i}}[\ell-1]]$ represents the time-variant RF fingerprint (of $A_{i}$), which captures the hardware fingerprints of the transmitter as well as the effects of the wireless channel between the transmitter and receiver. We model the received signal at the receiver by
\begin{equation} \label{io_eqn1}
    \mathbf{s}_R = \mathbf{H}_{R}^{A_{i}}\mathbf{x} + \mathbf{g},
\end{equation}
where $\mathbf{H}_R^{A_{i}} = \text{diag}(\mathbf{h}_{R}^{A_{i}}[0],\cdots,\mathbf{h}_{R}^{A_{i}}[\ell-1]) \in \mathbb{C}^{\ell \times \ell}$ and, $\mathbf{g} \in \mathbb{C}^{\ell}$ represents complex additive white Gaussian noise (AWGN). In addition, we consider an eavesdropper that will intercept the transmission of $\mathbf{x}$ through $\mathbf{h}_{E}^{A_{i}} \in \mathbb{C}^{\ell}$, where $\mathbf{h}_{E}^{A_{i}} \in [h_{E}^{A_{i}}[0], \cdots, h_{E}^{A_{i}}[\ell-1]]$ also captures the hardware fingerprints of $A_{i}$ as well as the effects of the wireless channel between the transmitter and the eavesdropper. We assume that the receiver and eavesdropper each has knowledge of their individual channel distribution between itself and the transmitter ($\mathbf{h}_{R}^{A_{i}}$ and $\mathbf{h}_{E}^{A_{i}}$, respectively), which is estimated from the pilot symbols of the received signal \cite{channelest2}. Specifically, both the eavesdropper and receiver, aware of the pilot symbols from the preamble, which is used to perform RFF, are able to recover the channel matrix through least squares channel estimation.

Due to the presence of the eavesdropper, the transmitter attempts to prevent fingerprinting at any unintended receivers by injecting a perturbation $\pmb{\delta} \in \mathbb{C}^{\ell}$ into $\mathbf{x}$ before transmission. Thus, the true signal transmitted by the transmitter is $\mathbf{x} + \pmb{\delta}$ and the eavesdropper receives 
\begin{equation}
    \mathbf{s}_{E} = \mathbf{H}_{E}^{A_{i}}(\mathbf{x} + \pmb{\delta}) + \mathbf{g}, 
\end{equation}
where $\mathbf{H}_{E}^{A_{i}} = \text{diag}(\mathbf{h}_{E}^{A_{i}}[0],\cdots,\mathbf{h}_{E}^{A_{i}}[\ell-1])\in \mathbb{C}^{\ell \times \ell}$. We note that the diagonal structure of $\mathbf{H}_R^{A_i}$ and $\mathbf{H}_E^{A_i}$ is well-justified in our IEEE 802.11g experimental setup. The 802.11g standard employs OFDM with a cyclic prefix of sufficient length to eliminate inter-symbol interference, so that after CP removal and the FFT at each receiver, the effective channel becomes a per-subcarrier multiplicative factor, yielding $Y[k] = H[k]X[k] + G[k]$ for each subcarrier $k$. In matrix form, this corresponds precisely to our diagonal model. Since RF fingerprinting is performed on the preamble IQ samples used to estimate the per-subcarrier channel via least-squares, the diagonal representation directly captures the hardware impairments that constitute the RF fingerprint. This model is consistent with prior RFF work on IEEE 802.11g signals \cite{channelest2,DLRF1}. Moreover, the distinct channel distributions of the receiver and eavesdropper arise naturally from the use of two physically separate USRP hardware units, each with unique hardware imperfections such as IQ imbalance and carrier frequency offset. Although the two devices are co-located in our experiments, these hardware differences produce distinct channel realizations without any physical displacement of the eavesdropper, and each device independently estimates its own diagonal channel matrix from the received preamble pilot symbols via least squares. Although unknown at the eavesdropper, the transmitter may use $\boldsymbol{\delta} = \mathbf{0}$ in which case we represent the intercepted signal at the eavesdropper as $\mathbf{s}_{E}^{\prime}$. We further characterize the design of $\pmb{\delta}$ in Sec. \ref{sec:ee}. With the injected perturbation, the true signal collected at the receiver is 
\begin{equation}
    \widetilde{\mathbf{s}}_{R} = \mathbf{H}_{R}^{A_{i}}(\mathbf{x} + \pmb{\delta}) + \mathbf{g}, 
\end{equation}
where $\tilde{\mathbf{s}}_{R} = \mathbf{s}_{R}$ (from (\ref{io_eqn1})) when the transmitter is not actively perturbing its transmission and, thus, $\pmb{\delta} = \mathbf{0}$.

Since RF fingerprinting at both the receiver and eavesdropper are deep learning-based (as further discussed in Sec. \ref{clf_mod}), we map all complex signals to real values $\mathbb{C}^{\ell} \rightarrow \mathbb{R}^{2\ell}$ using the real and imaginary signal components for compatibility with real-valued neural networks. 

\subsection{Receiver Modeling} \label{clf_mod}

The objective of the deep learning (DL) model at the receiver is to fingerprint a received signal as belonging to either the set of trusted devices, $\mathcal{T}$, or the set of untrusted devices $\mathcal{U}$. Here, the receiver trains a classifier on instances of $\mathbf{s}_{R}$, representing the received signal's preamble, which can be used in place of the entire received payload to perform effective device RF fingerprinting without incurring unnecessary training overhead \cite{preamble1,preamble2}. 

Formally, we define $\mathcal{X}_{\text{tr}} = \{\mathbf{s}_{R}^{(i)}, p^{(i)}; i = 1, \cdots,N\}$ as the training dataset, where $p^{(i)} \in \{0, 1\}$ denotes if $\mathbf{s}_{R}^{(i)}$ was transmitted from an untrusted device ($p^{(i)} = 0$) or a trusted device ($p^{(i)} = 1$) and $N$ denotes the number of training samples. Using this dataset, at the receiver, we train a deep neural network (DNN), $\zeta(\cdot, \pmb{\theta}): \mathbb{R}^{2\ell} \rightarrow \mathbb{R}$, parameterized by $\pmb{\theta}$, to map the $i^{\text{th}}$ input sample, $\mathbf{s}_{R}^{(i)} \in \mathbb{R}^{2\ell}$, (i.e., the preamble of the received signal, which captures the hardware impairments of its transmitter) to the probability $p \in \mathbb{R}$ that the transmitter of the received signal is a trusted device. The DNN takes input $(\cdot)$, and outputs the probability that the input was broadcast from a trusted transmitter. Specifically, the trained DNN assigns each testing input $\mathbf{s}_{R} \in \mathbb{R}^{2\ell}$ a label denoted by $\hat{p}(\mathbf{s}_{R}, \pmb{\theta}) = \zeta(\mathbf{s}_{R}, \pmb{\theta})$, where $\zeta(\mathbf{s}_{R}, \pmb{\theta})$ is the predicted classification probability of $\mathbf{s}_{R}$ being transmitted from a trusted device. A threshold of 0.5 is used to map $\hat{p}(\mathbf{s}_{R}, \pmb{\theta})$ to a $0$ or $1$. During evaluation, we employ the dataset $\mathcal{X}_{\text{te}} = \{\mathbf{s}_{R}^{(j)}, p^{(j)}; j = 1, \cdots,K\}$, where $\mathcal{X}_{\text{tr}} \cap \mathcal{X}_{\text{te}} = \emptyset$ (i.e., no samples overlap in the training and evaluation datasets). 

At the receiver, we consider a fully connected neural network (FCNN) architecture. FCNNs consists of $L$ layers, indexed by $l$, where the $l^{\text{th}}$ layer is parameterized by its weights $\pmb{\theta}^{(l)} \in \mathbb{R}^{d_{l} \times d_{l-1}}$ and biases $\mathbf{b}^{(l)} \in \mathbb{R}^{d_{l}}$ that are estimated from the training data and $d_{l}$ and $d_{l-1}$ are hyperparameters denoting the number of units in the $l$ and $l-1$ layer, respectively. The output of the $l^{\text{th}}$ layer is given by $\mathbf{a}^{(l)} = \sigma(\mathbf{z}^{(l)})$, where $\sigma(\cdot)$ is the element-wise activation function, $\mathbf{z}^{(l)} = \pmb{\theta}^{(l)}\mathbf{a}^{(l-1)} + \mathbf{b}^{(l)}$, $\mathbf{a}^{(0)} = \mathbf{s}_{R}$, and $\mathbf{a}^{(L)} = \hat{p}(\mathbf{s}_{R}, \pmb{\theta})$. 
The RFF classifier is trained by minimizing the cross entropy loss function given by 
\begin{equation} \label{loss_fcn}
    \mathcal{L}(\mathbf{s}_{R}, p, \pmb{\theta}) = -\frac{1}{N}\sum_{i=1}^{N} p^{(i)} \text{log}(\hat{p}^{(i)})
\end{equation}
using stochastic gradient descent (SGD).

\subsection{Evading Eavesdroppers} \label{sec:ee}
Due to the broadcast nature of wireless networks, eavesdroppers can intercept and fingerprint transmitted signals using their own surrogate classifier. Despite the eavesdropper and transmitter operating with different classifiers, \cite{advattacktransf1,advattacktransf2}, a perturbation crafted on a particular classifier (e.g., the receiver's classifier) has strong transferability to other classifiers (i.e., the eavesdropper's classifier) trained to perform the same task (e.g., RFF) \cite{advattacktransf3}. Thus, in an effort to evade eavesdroppers, the transmitter, prior to broadcasting $\mathbf{x}$, injects an adversarial perturbation, $\pmb{\delta}$, which is specifically crafted to fool any potential eavesdroppers and prevent accurate RFF at unauthorized sources while simultaneously ensuring that the receiver is able to recover the signal and perform accurate RFF. 

At the transmitter, $\pmb{\delta}$ is crafted, with access to the classifier parameters of the receiver and an approximation of $\mathbf{s}_{R}$ (obtained by modeling the distributions of $\mathbf{H}_R^{A_{i}}$ and $\mathbf{g}$ at the transmitters), to find an optimal solution to 
\begin{subequations} \label{adv_opt:all-lines}
\begin{align}
    \underset{\pmb{\delta}}{\text{min}} \quad &  ||\pmb{\delta}||_{p} \label{adv_opt:line_1} \\
    \text{s. t.} \quad &  \hspace{0.5mm} \hat{p}(\mathbf{s}_{R}, \pmb{\theta}) \neq \hat{p}(\tilde{\mathbf{s}}_{R}, \pmb{\theta}), \label{adv_opt:line_2} \\
    \quad & \zeta(\mathbf{s}_{R}, \boldsymbol{\theta}) = \zeta(g_{\pmb{\theta}_{g}}(h_{\pmb{\theta}_{h}}(\widetilde{\mathbf{s}}_{R})), \boldsymbol{\theta}), \label{adv_opt:line_3} \\
    \quad & ||\pmb{\delta}||_{2}^{2} \leq P_{T}, \label{adv_opt:line_4} \\ 
     \quad & \tilde{\mathbf{s}}_{R}  \in \mathbb{R}^{2\ell} \label{adv_opt:line5},
\end{align}
\end{subequations} 
where (\ref{adv_opt:line_1}) denotes the weakest $l_{p}$-bounded perturbation, (\ref{adv_opt:line_2}) aims to generate a transferable perturbation, (\ref{adv_opt:line_3}) ensures that the receiver can effectively perform RFF (more specifically described in (\ref{mse_loss}) below), (\ref{adv_opt:line_4}) constrains the power budget $ P_{T}$, and (\ref{adv_opt:line5}) ensures that the perturbation remains in the appropriate dimension. 


Due to its non-linear nature, a solution to \ref{adv_opt:all-lines} is not guaranteed to exist in which cases a misclassification may never be induced. Moreover, solving (\ref{adv_opt:all-lines}) is difficult in practice due to its non-linear nature. Therefore, we approximate a solution to (\ref{adv_opt:all-lines}) using first-order gradient-based adversarial perturbations because they are computationally efficient methods to quickly embed a perturbation in the signal prior to transmission. Note that optimization-based perturbations \cite{cwattack,dfattack} can also be used to craft adversarial attacks. However, unlike gradient-based attacks, optimization-based perturbations are generally less transferable to surrogate models due to being overly tailored to the decision boundary of the target model, allowing an eavesdropper to fingerprint a device encoded with an optimization-based attack by merely using a surrogate model. Furthermore, optimization-based perturbations require significantly higher computational complexity, leading to slower transmissions and overall communications. In contrast, one-step and multi-step gradient-based attacks that are extensively used in adversarial wireless communication settings provide a computationally efficient method to generate adversarial perturbations and are more likely to result in transferable perturbations on surrogate classifiers, making them a realistic and attractive scheme for adversarially encoded transmissions \cite{rffeavesdropper}. Thus, we consider the fast gradient sign method (FGSM), a \emph{single-step perturbation} \cite{fgsm}, and projected gradient descent (PGD), an \emph{iterative perturbation} \cite{pgd} injected by the transmitter as approximate solutions to (\ref{adv_opt:all-lines}).


The goal of FGSM is to take a single step, using its total power budget $P_{T}$, in the direction that maximizes the classifier's loss function. For an $l_{2}$ bounded perturbation, this is given by 
\begin{equation}
\pmb{\delta} = \sqrt{P_{T}} \frac {\nabla_{\mathbf{s}_{R}} \mathcal{L}(\mathbf{s}_{R}, p, \pmb{\theta})} {||\nabla_{\mathbf{s}_{R}} \mathcal{L}(\mathbf{s}_{R}, p, \pmb{\theta})||_{2}}, \label{l2_fgsm}
\end{equation}
and, for an $l_{\infty}$-bounded perturbation is given by
\begin{equation} \label{linf_fgsm}
    \pmb{\delta} = {\sqrt{P_{T}}} \cdot \text{sign}(\nabla_{\mathbf{s}_{R}} \mathcal{L}(\mathbf{s}_{R}, p, \pmb{\theta})), 
\end{equation}
where, in the $l_{\infty}$-bounded perturbation, each feature of $\mathbf{s}_{R} \in \mathbb{R}^{2\ell}$ is perturbed by $\sqrt{P_T}$ in the direction of the sign of the loss function's gradient. We emphasize that (\ref{l2_fgsm}) and (\ref{linf_fgsm}) are \emph{distinct} formulations: the $l_2$-bounded attack normalizes the full gradient vector to have $l_2$-norm $\sqrt{P_T}$, while the $l_\infty$-bounded attack applies the element-wise sign function and scales by $\sqrt{P_T}$, so that each element is bounded in magnitude by $\sqrt{P_T}$ (satisfying $\|\pmb{\delta}\|_\infty \leq \sqrt{P_T}$). Both scalings are consistent with the power budget constraint in (\ref{adv_opt:line_4}).


PGD is an iterative extension of FGSM that repeatedly steps in the gradient direction over multiple iterations, using a fraction of its power budget $\beta = 2\sqrt{P_T} / T$ at each iteration, where $T$ is the total number of iterations. Specifically, initializing $\pmb{\delta}^{(0)} = \mathbf{0}$, the $l_{2}$-bounded PGD perturbation on iteration $t+1$ is given by
\begin{equation} \label{l2_pgd}
    \pmb{\delta}^{(t+1)} = \text{Proj}_{B_{2}(0, \sqrt{P_{T}})} \bigg(\pmb{\delta}^{(t)} + \beta \frac {\nabla_{\widetilde{\mathbf{s}}_{R}^{(t)}} \mathcal{L}(\widetilde{\mathbf{s}}_{R}^{(t)}, p, \pmb{\theta})} {||\nabla_{\widetilde{\mathbf{s}}_{R}^{(t)}} \mathcal{L}(\widetilde{\mathbf{s}}_{R}^{(t)}, p, \pmb{\theta})||_{2}} \bigg), 
\end{equation}
where $\pmb{\delta}^{(T)} = \pmb{\delta}$, $\widetilde{\mathbf{s}}_{R}^{(t)} = \mathbf{s}_{R} + \pmb{\delta}^{(t)}$, and 
\begin{equation}
    \text{Proj}_{B_{2}(0, \sqrt{P_{T}})}(\cdot) = \sqrt{P_{T}} \frac{\cdot}{\text{max}(||\cdot||, \sqrt{P_{t}})}
\end{equation}
ensures the perturbation remains inside an $l_{2}$-ball of radius $\sqrt{P_{T}}$ around $\mathbf{s}_{R}$. Similarly, for an $l_{\infty}$-bounded PGD attack, the perturbation at iteration $t + 1$, using $\beta = \sqrt{P_{T}} / T$ and $\pmb{\delta}^{(0)} = \mathbf{0}$, is given by
\begin{equation} \label{linf_pgd}
    \pmb{\delta}^{(t+1)} = \text{Proj}_{B_{\infty}(0, \sqrt{P_{T}})} \bigg(\pmb{\delta}^{(t)} + \beta \text{sgn}(\nabla_{\widetilde{\mathbf{s}}_{R}^{(t)}} \mathcal{L}(\widetilde{\mathbf{s}}_{R}^{(t)}, p, \pmb{\theta})) \bigg), 
\end{equation}
where $\pmb{\delta}^{(T)} = \pmb{\delta}$, $\widetilde{\mathbf{s}}_{R}^{(t)} = \mathbf{s}_{R} + \pmb{\delta}^{(t)}$, and 
\begin{equation}
    \text{Proj}_{B_{\infty}(0, \sqrt{P_{T}})}(\cdot) = \text{clip}(\cdot, \{-\sqrt{P_{T}}, \sqrt{P_{T}}\})
\end{equation}
ensures the element-wise perturbation remains less than $\sqrt{P_{T}}$.

The transmitter injects $\pmb{\delta}$ into its signal prior to broadcasting it using either FGSM or PGD. As a result of the transferability of adversarial perturbations, any eavesdropper that intercepts the signal and attempts to perform RFF will classify the signal erroneously. The receiver, on the other hand, is fully aware of the presence of the perturbation in its received signal and is thus able to filter $\pmb{\delta}$ and retain its RFF performance despite the presence of the perturbation. 

Due to the effectiveness of Denoising Autoencoders (DAE) in our considered threat model \cite{MIMODAE,DAE2}, the receiver employs a DAE to reconstruct $\mathbf{s}_{R}$ from $\widetilde{\mathbf{s}}_{R}$. Specifically, after observing a received transmission, $\widetilde{\mathbf{s}}_{R} \in \mathbb{R}^{2\ell}$, where $P(\widetilde{\mathbf{S}}_{R} |\mathbf{S}_{R} = \mathbf{s}_{R})$ denotes the conditional distribution of a perturbed signal given its clean counterpart, an encoder defined as $Q_{\pmb{\theta}_Q}(\cdot):\mathbb{R}^{2\ell} \rightarrow \mathbb{R}^{q}$ ($q < 2\ell$) is trained to map $\widetilde{\mathbf{s}}_{R}$ into a lower dimensional feature representation. In the reconstruction process, a decoder $G_{\pmb{\theta}_G}(\cdot):\mathbb{R}^{q} \rightarrow \mathbb{R}^{2\ell}$ learns to map the lower dimensional feature representation to an approximation of the input without the perturbation. The encoder and decoder are jointly trained to learn the distribution of $P(\mathbf{S}_{R} |\widetilde{\mathbf{S}}_{R} = \widetilde{\mathbf{s}}_{R})$ by minimizing the loss of the sample between the perturbed signal $\widetilde{\mathbf{s}}_{R}$ and corresponding unperturbed signal $\mathbf{s}_{R}$. Here, we minimize the mean squared error loss, which is given by 
\begin{equation}
    \mathcal{L}_{\text{MSE}} = \frac{1}{N} \sum_{i=1}^{N} \frac{1}{2\ell} \sum_{k=1}^{2\ell} (\mathbf{s}_{R}^{i,k} - G_{\pmb{\theta}_{G}}(Q_{\pmb{\theta}_{Q}}(\widetilde{\mathbf{s}}_{R}^{i,k})) )^{2}. \label{mse_loss}
\end{equation}
Using SGD, the parameters $\pmb{\theta}_Q$ and $\pmb{\theta}_G$ are tuned to reach a local minimum that sufficiently reduces (\ref{mse_loss}). In addition to $\{\widetilde{\mathbf{s}}_{R}^{(i)}, \mathbf{s}_{R}^{(i)}; i = 1, \cdots,N\}$, we also include $\{\mathbf{s}_{R}^{(i)}, \mathbf{s}_{R}^{(i)}; i = 1, \cdots,N\}$ as training samples so that the DAE is capable of mapping benign transmissions (i.e., received signals that do not contain adversarial perturbations from the transmitter) to a version of itself.

Although the transmitter and intended receiver are aware that the transmissions may be embedded with deliberate adversarial perturbations, the receiver solely observes the potentially corrupted transmission without the ability to directly isolate the perturbation. Thus, the transmitter may inject any $\boldsymbol{\delta}$ into its broadcast signal. As a result, the receiver, which will never know the true value of $\boldsymbol{\delta}$ used by the transmitter, must be prepared to accurately fingerprint a corresponding device regardless of the value of $\boldsymbol{\delta}$. To ensure this is possible at the receiver, the received signal is first propagated through a DAE in order to filter out any potentially additive perturbation. The denoised version of the signal is then used for RFF. The DAE is trained on perturbation-free ($\boldsymbol{\delta}= 0$) RF signatures as well as perturbations of varying potencies in order to reconstruct the unperturbed approximation of the signal regardless of the power of $\boldsymbol{\delta}$.

During deployment, the denoised signal $G_{\pmb{\theta}_{G}}(Q_{\pmb{\theta}_{Q}}(\widetilde{\mathbf{s}}_{R}^{i,j}))$ is forward propagated through the classifier $\zeta(\cdot, \pmb{\theta}): \mathbb{R}^{2\ell} \rightarrow \mathbb{R}$, yielding $\hat{p}(\mathbf{s}_{R}, \pmb{\theta}) \in \mathbb{R}$, which denotes the estimated probability with which the received signal was transmitted from a trusted device. As a result of this design, the transmission severely degrades the eavesdropper's ability to gain sensitive information, due to the transferability property of adversarial perturbations, while the receiver can effectively filter the injected noise to obtain an accurate representation of the original transmission, thus allowing for secure communication between the transmitter and receiver.

In order to ensure an effective adversarial attack, the transmitter strategically injects $\boldsymbol{\delta}$ into the preamble of the broadcast signal. The receiver and potential eavesdropper exploit the provided short training field to extract parameters including symbol timing and carrier frequency offset enabling detection of the start of the signal frame and symbol boundary. After proper synchronization, the receiver and eavesdropper compare the received pilot symbols with the target symbols provided by the long training field to estimate the channel at each subcarrier \cite{channelest2}. As a result of the synchronization, the perturbation impacts the received symbols, leading to misclassifcation.

Moreover, note that the additive perturbation does not impact synchronization at the receiver. Synchronization is performed via correlation with a known and deterministic preamble. As a result, bounded additive perturbations behave similarly to additional noise and do not eliminate correlation peaks unless their power approaches or exceeds that of the signal. Since our perturbations are strictly power-constrained ($\text{PSR} < 0$ dB), they do not disrupt frame detection or symbol boundary estimation. This insight has been leveraged by prior work as well to encode adversarial perturbations into transmitted signals that inject perturbations while preserving decodability \cite{evs1,evs3,evs4,evs5,amc_eaves,rffeavesdropper}.

\subsection{Resilience Analysis of Eavesdropper} \label{thm} 

Given the design presented in Sec. \ref{sec:ee} for communication between a transmitter and receiver, we now shift our focus to the eavesdropper. Here, we develop, and theoretically demonstrate, a stealthy approach for the eavesdropper to perform effective RFF despite the covertness between the transmitter and receiver. 

Let us first define the threat model of the eavesdropper. We evaluate the potency of the transmitter's injected perturbation under two distinct environments: (i) Environment A: Partial Receiver Privacy in which the eavesdropper has partial but incomplete knowledge about the classification architecture at the receiver and (ii) Environment B: Complete Receiver Privacy in which the eavesdropper has no knowledge of the classification architecture at the receiver. More specifically, in Environment A, the eavesdropper has knowledge of the receiver's DNN classifier architecture but is blind to its parameters as well as to the DAE whereas, in Environment B, the eavesdropper is blind to both the receiver's DNN architecture and the DAE. We omit from consideration the unrealistic scenario in which the eavesdropper has complete knowledge of the receiver because (i) it is rare (if possible) for an eavesdropper to have perfect knowledge about the signal processing module of the intended receiver in practice \cite{real_world} and (ii) in such a scenario, the eavesdropper can directly adopt the mitigation strategy at the receiver.

From the eavesdropper's perspective, the goal is to fingerprint $\mathbf{s}_{E}$ as either a trusted or untrusted device, regardless of any perturbations in the intercepted signal. Existing methods have shown that the transferability of $\pmb{\delta}$ is limited outside of the time domain \cite{adv_trf}, but such methods lack robustness, since the baseline performance on unperturbed signals outside of the time-domain is typically lower compared to using time domain IQ samples \cite{diff_dom}. To address this, we propose to use the short time Fourier transform (STFT) to divide $\mathbf{s}_{E}$ into equal sized intervals and capture the frequency spectrum of each segment, thus preserving the time-domain information on which DNN-based classifiers are known to be most effective. The STFT of the received signal at the eavesdropper is defined as
\begin{equation}
    S_{E}^{(n,f)} = \sum_{m=-\infty}^{\infty} s_{E}[m] w[m-n] e^{-j(2{\pi}f/M)m}, 
\end{equation}  
where $S_{E}^{(n,f)}$ represents the time-frequency component at time index $n$ and frequency index $f$, $w[m-n]$ is the Gaussian window function centered at $m = n$, $e^{-j(2{\pi}f/M)m}$ is the exponential basis function, and $M$ is the total number of discrete Fourier transform bins. Moreover, we use $\mathbf{S}_{E} = \mathcal{S}[\mathbf{s}_{E}] \in \mathbb{R}^{T \times F}$ to represent the STFT-based spectrogram of $\mathbf{s}_{E}$, where $F$ denotes the number of frequency bins for each segment of $\mathbf{s}_E$ and $T$ is the number of segments of $\mathbf{s}_E$. 

The eavesdropper uses disparate datasets from the receiver, since channel conditions would prevent the eavesdropper from capturing the receiver's exact received signals, denoted by $\mathcal{X}_{\text{tr}}^{E} = \{\mathbf{S}_{E}^{(i)}, p^{(i)}; i = 1, \cdots,N\}$ and $\mathcal{X}_{\text{te}}^{E} = \{\mathbf{S}_{E}^{(j)}, p^{(j)}; j = 1, \cdots,K\}$ to train its own DNN, $\xi(\cdot, \mathbf{w}): \mathbb{R}^{TF} \rightarrow \mathbb{R}$ (where we reshape the spectrogram to maintain compatibility with DNNs), parameterized by $\mathbf{w}$, to predict the probability, $\hat{p}(\mathbf{S}_{E}, \mathbf{w}) = \xi(\mathbf{S}_{E}, \mathbf{w})$, that the transmitter of the received signal is a trusted device. 
Not only do we show that baseline performance of RFF is maintained on this time-frequency signal representation, in comparison to merely using IQ signals, but we also show that the eavesdropper can perform accurate classification on perturbed RFF signals, owing to feature importance being spread across time and frequency rather than merely being concentrated in the time domain. 



We quantify feature importance using SHapley Additive exPlanations (SHAP) \cite{shap}, which quantifies the contribution of each input feature to the DNN's prediction. Specifically, the SHAP value of $S_{E}^{(n,f)}$ is defined by 
\begin{equation}
    \phi_{S_{E}^{(n,f)}} = \nonumber 
\end{equation}
\begin{equation}
    \sum_{\mathcal{P}\subseteq \mathcal{Q} \backslash \{S_{E}^{(i)}\}} \frac{|\mathcal{P}|(|\mathcal{Q}|-|\mathcal{P}|-1)!}{|\mathcal{Q}|!} (\xi(\mathcal{P} \cup \{S_{E}^{(n,f)}\}, \boldsymbol{\theta}) - \xi(\mathcal{P}, \boldsymbol{\theta})), \label{shap_val_formula} 
\end{equation} 
where $\mathcal{P}$ is a feature subset of $\mathcal{Q}$, which contains the set of all features of $\mathcal{S}[\mathbf{s}_{E}]$, 
$\zeta(\mathcal{P} \cup \{S_{E}^{(n,f)}\}, \boldsymbol{\theta}) - \zeta(\mathcal{P}, \boldsymbol{\theta})$ is the difference between the predicted output from the model with and without $S_{E}^{(n,f)}$, which is summed across all possible feature subsets $\mathcal{P}$, and $\frac{|\mathcal{P}|(|\mathcal{Q}|-|\mathcal{P}|-1)!}{|\mathcal{Q}|!}$ is a normalization constant. 


Given these definitions, we will now analyze the impact of $\boldsymbol{\delta}$ on $\mathbf{S}_{E}$ and show that the eavesdropper can effectively perform RFF in its presence. We will employ the following assumptions in our analysis, where each assumption is verified in Sec. \ref{feat_imp_sec}: 


\noindent \textbf{Assumption 1:} When $\boldsymbol{\psi} = \mathbf{H}_E\boldsymbol{\delta}$ is dispersed across the signal (e.g., when $\boldsymbol{\delta}$ is $l_{2}$ bounded), $\mathcal{S}[\boldsymbol{\psi}] \approx 0 \hspace{1mm} \forall \hspace{1mm} f \notin (f_\psi, f_\psi + \Delta_\psi)$ and $f_\psi >> f_b$, where $f_\psi + \Delta_\psi$ and $f_{b}$ denotes the bandwidth of $\boldsymbol{\boldsymbol{\psi}}$ and  the high-frequency cutoff of $\mathbf{S}_{E}$, respectively. Thus, for $l_{2}$-bounded attacks, only the high frequency components of $\mathbf{S}_{E}$ are impacted. 


\noindent \textbf{Assumption 2:} When $\boldsymbol{\psi}$ consists of uniformly applied high-magnitude perturbations across the signal (e.g., when $\boldsymbol{\delta}$ is $l_{\infty}$ bounded), $\mathcal{S}[\boldsymbol{\psi}] \approx 0 \hspace{1mm} \forall \hspace{1mm} f \notin (f_{\psi}, f_\psi + \Delta_\psi )$ and $f_{\psi} +\Delta_\psi << f_{a}$, where $f_{a}$ denotes the low-frequency cutoff of $\mathbf{S}_{E}$ . Thus, for $l_{\infty}$-bounded attacks only low frequency components of $\mathbf{S}_{E}$ are impacted. 


\noindent \textbf{Assumption 3:} When analyzing $\mathbf{S}_{E}$, $\phi_{S_{E}^{(n,f^{*})}} >> \phi_{S_{E}^{(n,\bar{f})}}$, where $f^{*} \in [f_{a},f_{b}]$, and $\bar{f} \in (-\infty, f_{a}) \cup (f_{b}, \infty)$. Thus, high and low frequency components are not pivotal features for prediction in comparison to middle frequency indices. 




Given Assumptions 1 -- 3, and without loss of generality on the $l_{p}$-bound of $\boldsymbol{\delta}$, we present the following theorem to analytically characterize the effect of $\boldsymbol{\delta}$ on the time-frequency features of $\mathbf{s}_{E}$.  
\begin{theorem} \label{thm1}
Let $\psi = \mathbf{H}_E\pmb{\delta}$ denote the time-domain perturbation observed at the eavesdropper, where $\|\pmb{\delta}\|_2^2 \leq P_T$, and let $w[\cdot]$ be the Gaussian STFT window with standard deviation $\sigma$, satisfying the spectral tail bound $|W(\hat f)| \leq Ke^{-\hat f^2/(2\sigma^2)}$ for some constant $K > 0$. Suppose $\Psi(\hat f) \approx 0$ for all $\hat f$ outside a band $(f_\psi, f_\psi+\Delta_\psi)$ (from Assumption~1 or Assumption~2), and let $\kappa > 0$ be such that $|\hat f - f| \geq \kappa\sigma$ for every $\hat f$ in this band and every $f \in (f_a, f_b)$. Then, for every $f \in (f_a, f_b)$ and every time index $n$, the PSD of the received signal at the eavesdropper satisfies
\begin{equation}
\Big|\,|S_E^{(n,f)}|^2 - |S_E^{\prime(n,f)}|^2\,\Big| \leq\  K^2 P_T \|\mathbf{H}_E\|_2^2 \Delta_{\psi}^{2}  e^{-\kappa^2} +\nonumber
\end{equation}
\begin{equation}  \label{thm1_bound}
2K\sqrt{P_T}\,\|\mathbf{H}_E\|_2\, e^{-\kappa^2/2}\Big(|Y^{(n,f)}| + |G^{(n,f)}|\Big),
\end{equation}
where $S_E^{\prime(n,f)}$ denotes the STFT of the received signal when $\pmb{\delta} =\mathbf{0}$. In particular, the right-hand side of \eqref{thm1_bound} is finite for every finite $\kappa$ and tends to zero as $\kappa \to \infty$, so that the PSD of the perturbed signal converges to the PSD of the unperturbed signal in the mid-band $(f_a, f_b)$ as the frequency separation between the perturbation and the mid-band grows.
\end{theorem}


\begin{proof} See Appendix A. \end{proof}

Theorem \ref{thm1} shows that a transferable perturbation constructed at the transmitter has a negligible effect on the spectrogram of the received signal at the eavesdropper. 
Therefore, the PSD of the received signal at the eavesdropper, $\mathbf{s}_{E}$, when $\boldsymbol{\delta} \neq \mathbf{0}$ is approximately equal to the PSD when $\boldsymbol{\delta} = \mathbf{0}$. Thus, the spectrogram-based RFF classifier at the eavesdropper is expected to retain high classification performance when a signal containing adversarial perturbations is intercepted by the eavesdropper. 

Furthermore, Theorem \ref{thm1} provides direct guidance for eavesdropper architecture design. Since transferable time-domain perturbations induce negligible changes in the PSD of the received signal, robust RFF at the eavesdropper should operate on magnitude-based STFT representations rather than raw IQ samples. In conjunction with Assumptions 1 -- 3, this result further indicates that discriminative information is concentrated in mid-band frequency components, while perturbation energy is largely confined to low or high-frequency regions depending on the norm constraint. In practice, we observe that this invariance holds for power-constrained perturbations (as required in (\ref{adv_opt:line_4})), beyond which PSD deviations begin to mask the preamble itself. To make the bound in Theorem~1 quantitatively precise, we note that for $f \in (f_a, f_b)$, the bound in Theorem 1 evaluates to approximately $e^{-4.5} \approx 0.011$ at $\kappa=3$. For our experimental STFT parameters ($\sigma = 7$, $L = 50$, hop $= 20$), a frequency separation of $\kappa = 3$ standard deviations ($3\sigma = 21$ frequency bins) between the perturbation band and the mid-band signal region yields $\epsilon_\kappa \approx e^{-4.5} \approx 0.011$, i.e., less than $1.1\%$ of the peak window response. This illustrates one of many practical design guideline: STFT window parameters should be chosen so that the frequency separation between the perturbation energy and the fingerprint-bearing mid-band components is at least $3\sigma$. Thus, Theorem 1 jointly verifies the feasibility of our framework and defines a practical deployment regime for STFT-based eavesdropping.

\section{Performance Evaluation} \label{results}

In this section, we conduct an empirical evaluation of our proposed eavesdropper-based RFF framework. First, we discuss our empirical setup including our employed dataset, classifier architectures, and experimental parameters (Sec. \ref{exp_setup}). Next, we evaluate the baseline RFF performance of the receiver and eavesdropper (Sec. \ref{baseline_perf}) and analyze the feature importance across frequencies to verify Assumptions 1 -- 3 (Sec. \ref{feat_imp_sec}). Finally, we evaluate the efficacy of our framework on multiple perturbation designs and potencies and report our results in comparison to several considered baseline approaches (Sec. \ref{fgsm_res} and Sec. \ref{pgd_res}).


\subsection{Experimental Setup} \label{exp_setup}
We employ a subset of the RF fingerprinting dataset in \cite{datasetpaper} for our empirical evaluation. Here, we consider a total of $|\mathcal{A}| = 30$ devices, where we select $|\mathcal{T}| = 5$ trusted transmitters and $|\mathcal{U}| = 25$ untrusted transmitters (10 transmitters are unseen and untrusted). In our empirical setup, each transmitter is placed one meter apart in a grid-like fashion, surrounding a single receiver (Universal Software Radio Peripheral (USRP) N210) and we simulate an identical eavesdropper placed next to the receiver, where the receiver and eavesdropper record WiFi signals sent over the IEEE 802.11g Channel 11. The adjacent placement of the eavesdropper resembles the worst case scenario in which from the eavesdropper perspective, the adversarial potency is most detrimental, while negligibly influencing the channel conditions of the receiver due to the passive nature of the eavesdropper.\footnote{Prior work has noted that the effect of the channel distribution reduces the effectiveness of adversarial perturbations \cite{channeleffectpert1,channeleffectpert2,channeleffectpert3}.} We assume that the eavesdropper does not have any information about the observations used by the receiver for training, but has obtained its own observations of signals collected from channel conditions distinct from those observed by the receiver, corresponding to the set of trusted and untrusted transmitters, which are used for training at the eavesdropper. The receiver and eavesdropper each use $28334$ training signals, with 17450 signals received from trusted devices and 10884 signals received from untrusted devices, where no signal between them are the same, and each signal contains $\ell = 256$ samples. We account for dynamically changing conditions by training the eavesdropper and receiver on waveforms collected over four separate days and testing it on waveforms collected on a distinct fifth day, allowing the models to generalize to varying environments without additional overhead complexity.


In our empirical analysis, we use the perturbation-to-signal ratio (PSR) to quantify the perturbation magnitude. Formally, the PSR is given by 
\begin{equation}
     \text{PSR} \hspace{1mm} [\text{dB}] = \frac {\mathbb{E}[\|\pmb{\delta}\|_{2}^2]}  {\mathbb{E}[\|\mathbf{s}\|_2^2]}  \hspace{1mm} [\text{dB}] ,
\end{equation}
where $\mathbb{E}$ denotes the expected value and $\mathbf{s}$ denotes the corresponding signal including noise, $\mathbf{n}$, and excluding the adversarial perturbation $\boldsymbol{\delta}$. To embed a perturbation within the signature, a gradient based perturbation is crafted and downsampled to the same numerical precision (15-17 decimal digits) as the intended clean signal. In the transmitter’s perspective, the additional step in fooling potential eavesdroppers leads to the extra overhead cost due to crafting adversarial perturbations which can be adjusted based on the complexity of the surrogate model and the user defined power budget. Here, it is desirable to keep $\text{PSR} < 0$, which can result in a transferable adversarial perturbation, so that the perturbation (i) is not detectable by the eavesdropper, (ii) does not overpower and mask the received signal, and (iii) can be effectively filtered out at the receiver to retain its expected RFF performance. 

In terms of performance, we focus on classification accuracy as well as loss rather than the bit error rate (BER) or frame error rate (FER), because the BER and FER are not impacted by the proposed approach. Since RF fingerprinting is performed using preamble IQ samples, which are deterministic and do not contain payload data bits, and our adversarial perturbations are injected exclusively into these preamble samples, the IQ samples corresponding to payload symbols remain unmodified. Thus, the demodulated bitstream is unchanged, and the BER and FER remain unaffected.

At the receiver, we train a lightweight RFF FCNN (i.e., a low-parameter classifier for computational efficiency to allow the FCNN to be deployed on edge devices such as routers) consisting of 3 hidden layers with 256, 128, and 64 units, respectively, and each hidden layer applies a 50\% dropout rate during training. In addition, the DAE architecture that we employ at the receiver consists of 5 hidden layers with 256, 128, 64, 128, and 256 units, respectively, and each hidden layer applies a 50\% dropout rate during training. The DAE uses a PSR of $-20$ dB for training which was empirically determined to maximize both the (i) classification accuracy on instances of $\widetilde{\mathbf{s}}_{R}$ at the receiver and (ii) transferability of the perturbation to the eavesdropper.
At the eavesdropper, we model two distinct system knowledge levels (as described in Sec. \ref{thm}): Environment A and Environment B. In Environment A, the RFF architecture at the eavesdropper is identical to the receiver's classifier architecture (excluding the DAE). In Environment B, the eavesdropper's classifier consists of two hidden layers with 100 and 80 units, respectively. We show that each of the considered lightweight classifiers are able to obtain classification accuracy comparable to the benchmark reported in \cite{datasetpaper} in the absence of adversarial attacks. At the eavesdropper, we use a classifier architecture identical to the receiver except the input layer is $600 \times 1$ to accommodate the STFT features. The standard deviation and window length were set to 7 and 50, respectively, after sufficient experimentation. For full reproducibility, the exact Gaussian window used is $w[n] = e^{-n^2/(2\sigma^2)}$ with $\sigma = 7$ and $n \in \{0, 1, \ldots, L-1\}$ where $L = 50$. The hop length (stride) between successive windows is $h = 20$ samples, yielding an overlap of $L - h = 30$ samples between adjacent frames. No additional zero-padding is applied beyond the window boundary itself. The STFT converts an $\ell = 256$ IQ signal into an $F = 50 \times T = 12$ spectrogram, corresponding to the $600 \times 1$ input layer. As a result, the individual features of the STFT spectrogram may be characterized by $(n,f)$ (representing the $n^{\text{th}}$ time segment of the the $f^{\text{th}}$ frequency component of the spectrogram of the received signal), where each  fingerprint is decomposed into $T$ time windows and the frequency resolution is adjusted to contain $F$ bins. Further, $n = 0$ and $n = T-1$ represent the first and last partition of the signal, and $f = 0$ and $f = F-1$ represents the lowest and highest frequency component at the $n^{\text{th}}$ time segment, respectively.

\begin{figure}[t] 
	\centering
	\includegraphics[width=\halfwidth\columnwidth]{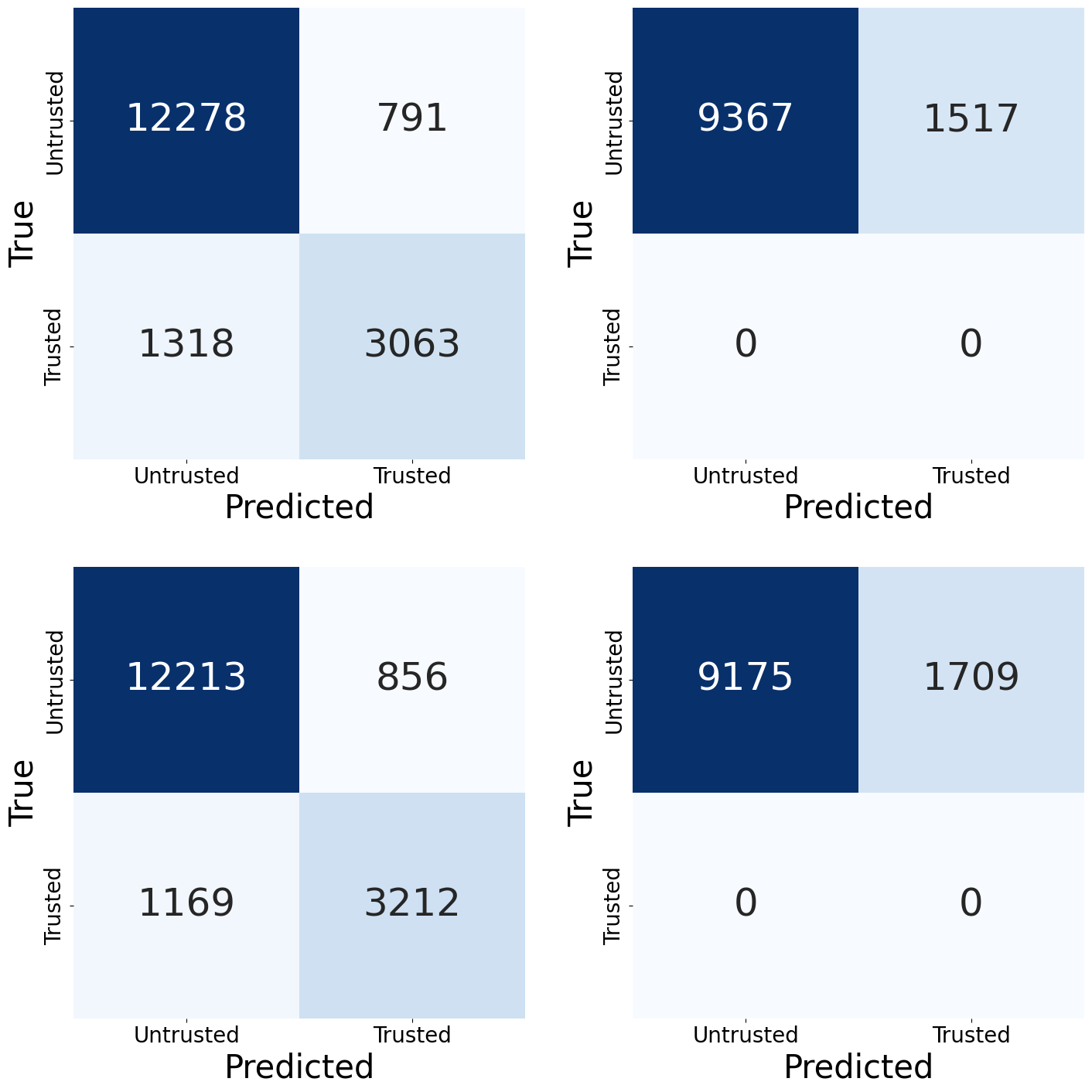}
	\caption{Confusion matrices showing baseline performance (i.e., with no adversarial attack injected by the transmitter) on signals received from seen and unseen transmitters (top left and top right, respectively) at the eavesdropper and seen and unseen transmitters (bottom left and bottom right, respectively) at the receiver.}
	\label{conf_matx}
\end{figure}

\begin{table}
    \caption{The accuracy, precision, recall, and F1-score of the receiver in each setting showing baseline performance on signals received from seen and unseen transmitters at the eavesdropper and receiver.}
    \centering
    \begin{tabular}{|c|c|c|c|c|}
        \hline 
        \textbf{Model} & \textbf{Acc.} & \textbf{Precision} & \textbf{Recall} & \textbf{F1-Score}\\
        \hline
         Seen Tx (Eavesdropper) & 0.8797 & 0.9395 & 0.9030 & 0.9209 \\
         Unseen Tx (Eavesdropper) & 0.8606 & 0.8606 & 1.0 & 0.9250 \\
         Seen Tx (Receiver) & 0.8840 & 0.9345 & 0.9126 & 0.9234 \\
         Unseen Tx (Receiver) & 0.8430 & 0.8430 & 1.0 & 0.9148 \\
         \hline 
    \end{tabular}
    \label{tab:results}
\end{table}

\subsection{RF Fingerprinting Performance} \label{baseline_perf}

We begin by analyzing the baseline performance of the eavesdropper and receiver on signals in which the transmitter did not inject an adversarial perturbation. Here, the eavesdropper and receiver achieve an accuracy of $87.91\%$ and $88.40\%$, respectively, demonstrating their strong RFF classification performance in terms of distinguishing trusted and untrusted devices. To provide a more complete picture of classification performance under class imbalance (5 trusted vs.\ 25 untrusted transmitters), we highlight that the confusion matrices in Fig.~\ref{conf_matx} provide a class-wise breakdown of performance, and, moreover, we calculate the precision, recall, and F1-score for both the eavesdropper and receiver under seen and unseen transmitter conditions in Table \ref{tab:results}. This performance is consistent with the classification performance achieved by \cite{datasetpaper} on their different day testing set (i.e., when the classifier was evaluated on RFF signatures collected on a different day than the training data), which is closely related to the testing set used in the evaluation of our framework, highlighting our method's ability to achieve high accuracy that is consistent with prior work.
However, each classifier must be able to effectively identify untrusted transmitters that were \emph{unseen} by the model during training. Thus, in an effort to characterize the performance of the eavesdropper and receiver on \emph{seen} and \emph{unseen} transmitters, we compute the accuracy on a disparate \emph{unseen} testing set. On an \emph{unseen} testing set (containing no signals from trusted devices), the eavesdropper and receiver attain accuracies of $86.06\%$ and $84.30\%$, respectively, thus showing their ability to identify untrusted devices not present in their training sets. We further show the performance of each setting in Fig. \ref{conf_matx}, where we see that our proposed eavesdropping framework achieves roughly the same performance as the receiver in authenticating both seen and unseen transmitters. 


\subsection{Feature Importance Analysis} \label{feat_imp_sec}

\begin{figure}[t] 
	\centering
	\includegraphics[width=\halfwidth\columnwidth]{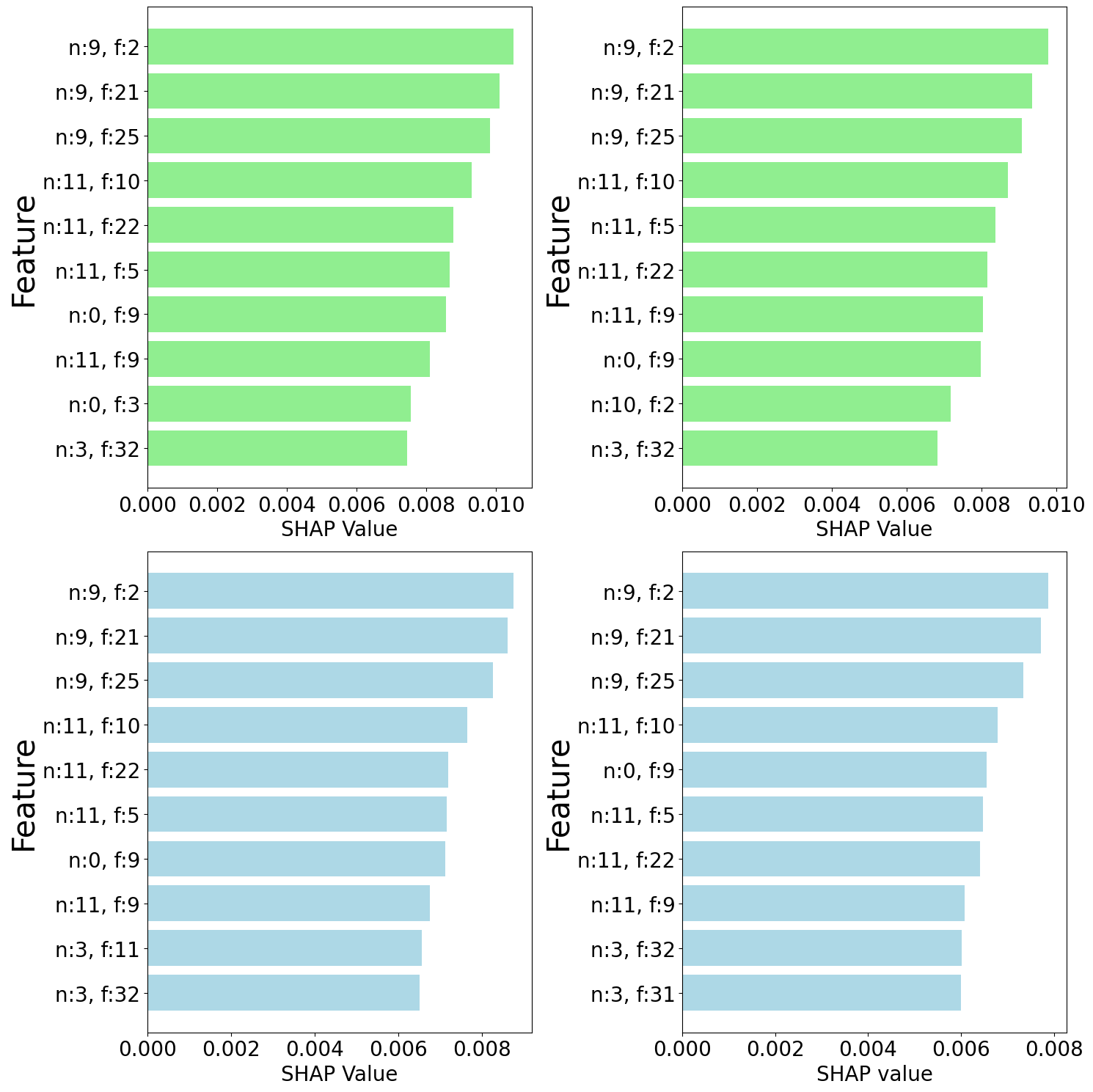}
	\caption{SHAP values, defined in (\ref{shap_val_formula}), corresponding to the time-frequency components of received signals perturbed with $l_2$ FGSM and $l_\infty$ FGSM (top left and top right) and $l_2$ and $l_\infty$ PGD (bottom left and bottom right) with $\text{PSR} = -20$ dB, where $n$ and $f$ represent the SHAP value at the $f^{\text{th}}$ frequency bin at the $n^{\text{th}}$ time segment. The features with the highest importance for RFF are generally middle frequencies, albeit with occasional high and low frequency components present depending on the norm bound, thus indicating that feature importance changes depending on the perturbation design.}
	\label{shap_vals}
\end{figure}

\begin{figure}[t] 
	\centering
	\includegraphics[width=\halfwidth\columnwidth]{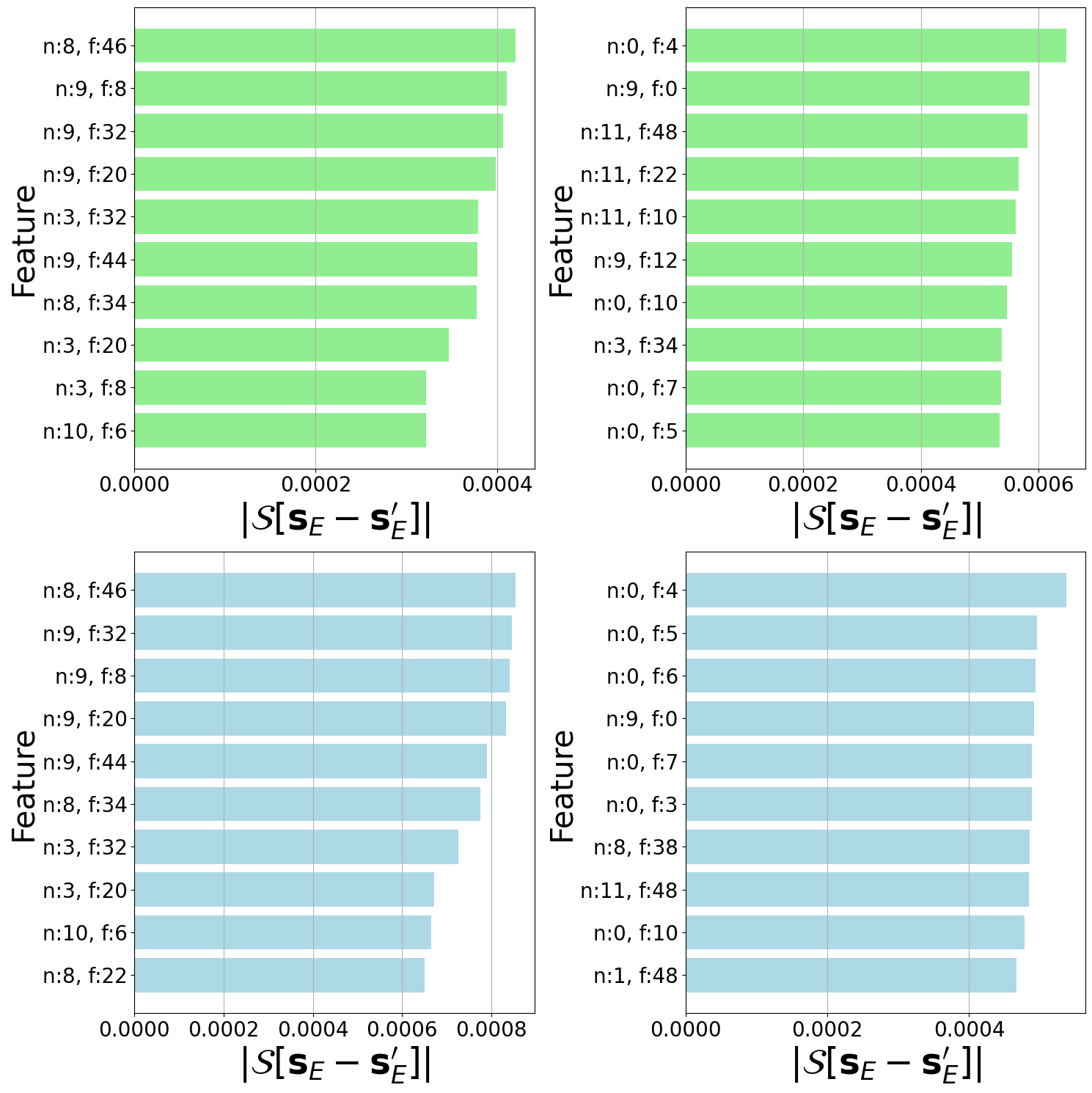}
	\caption{Here, we show the time-frequency features that are most affected by the added perturbation, calculated by taking the absolute difference of the STFT of the clean and perturbed signals. The top left and top right show the $l_{2}$- and $l_{\infty}$-bounded FGSM perturbation, and bottom left and bottom right show the $l_{2}$- and $l_{\infty}$-bounded PGD perturbation, where the transmitter crafts the perturbation at $\text{PSR} = -20$ dB. Here, $l_{2}$-bounded perturbations generally have a greater on impact high frequency components whereas $l_{\infty}$-bounded perturbations generally have a greater impact on low frequency components.}
	\label{feat_imp}
\end{figure}


We now perform a feature importance analysis of our proposed method to empirically corroborate Theorem 1 and confirm that Assumptions 1 -- 3 are valid. Here, we present our results in Environment A as we found that the feature analysis results are nearly identical in Environment B, so we omit them for brevity. Fig. \ref{shap_vals} shows our feature analysis, assigning SHAP values to the most significant features used by the eavesdropper's STFT classifier. From Fig. \ref{shap_vals}, we see that the STFT classifier primarily focuses on mid-range frequencies in adversarial environments indicating that both high frequency and low frequency components  have little influence on the classifier's prediction, thus confirming Assumption 3.

Moreover, to validate Assumption 1 and Assumption 2, we show the magnitude of the STFT features of the perturbation, $\boldsymbol{\delta}$, in Fig. \ref{feat_imp} for $l_{2}$ and $l_{\infty}$-bounded FGSM and PGD perturbations. From Fig. \ref{feat_imp}, we see that $l_{2}$-bounded perturbations largely impact high frequency components whereas $l_{\infty}$-bounded perturbations largely impact low frequency components. Along with Fig. \ref{shap_vals}, this shows that the features that are the most significantly impacted by adversarial perturbations, independent of the norm bound or perturbation algorithm, are not highly relevant in the classifier's prediction. This is consistent with Theorem 1 and demonstrates that applying the STFT to adversarially perturbed signals at the eavesdropper has negligible impact on the important features required for effective RFF at the eavesdropper. As a result, adversarial attacks injected by the transmitter are not expected to be highly transferable to the STFT features considered by the eavesdropper, thus allowing the eavesdropper to effectively operate in the presence of adversarial attacks with high performance. Note that empirical evaluations corroborating Assumptions 1 -- 3 were conducted at multiple PSR levels and yielded nearly identical results to the results shown at $\text{PSR} = -20$ dB and were thus omitted them for brevity.


\begin{figure}[t] 
	\centering
	\includegraphics[width=\halfwidth\columnwidth]{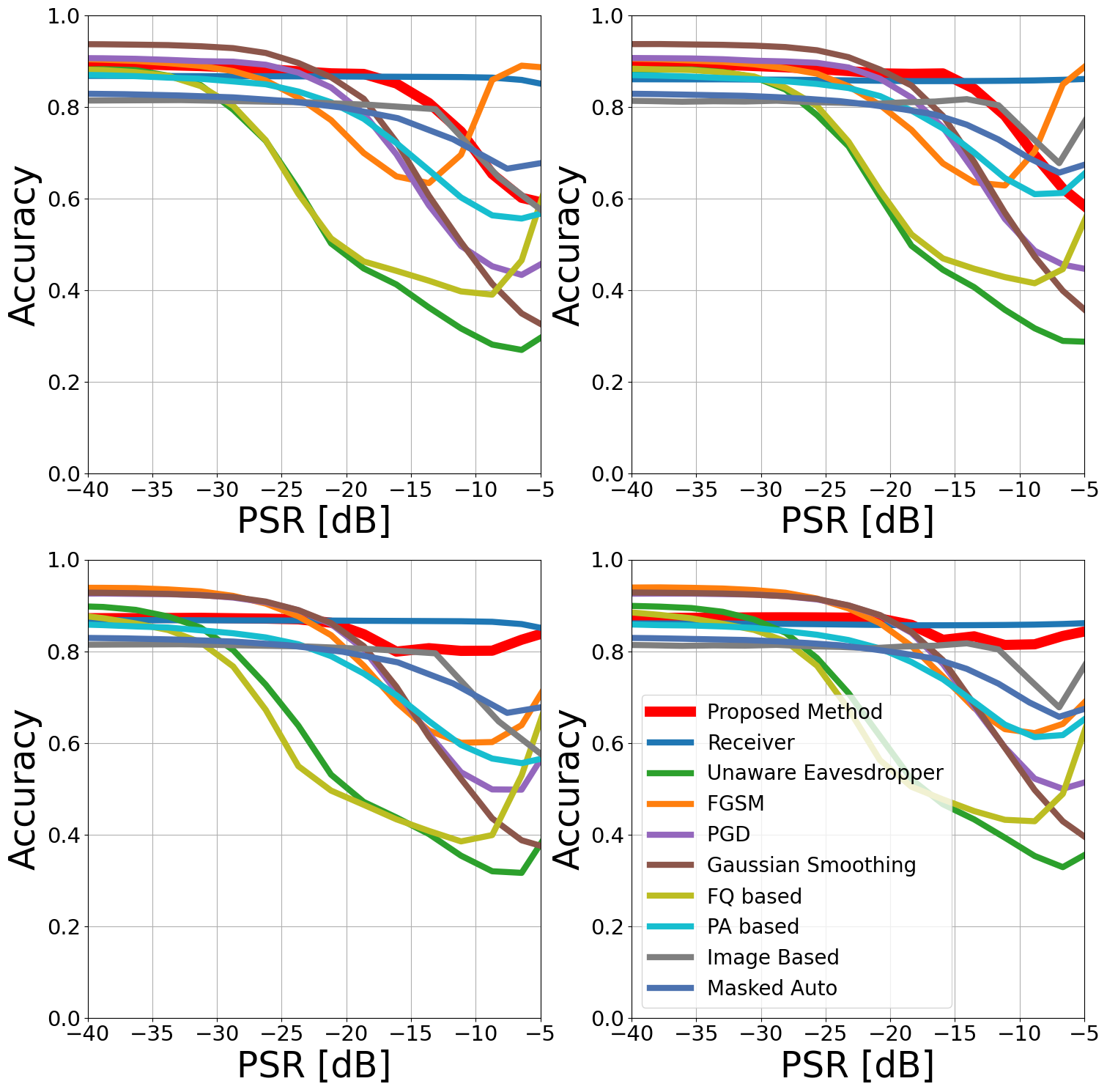}
	\caption{Eavesdropper's classification performance against FGSM perturbations injected by the transmitter in comparison to the receiver, unaware eavesdropper, and each considered baseline under $l_{2}$- and $l_{\infty}$-bounded perturbations in Environment A (top left and top right), and $l_{2}$- and $l_{\infty}$-bounded perturbation in Environment B (bottom left and bottom right). Under our proposed method, the eavesdropper outperforms each considered baseline and approaches the performance of the receiver in certain instances.}
	\label{fgsm_res_im}
\end{figure}


\subsection{Eavesdropper Resilience Against Single-Step Perturbations} \label{fgsm_res}

Here, we analyze the resilience of our eavesdropper's RFF approach against single-step FGSM perturbations. Although specific methods do not currently exist to strengthen an eavesdropper's robustness against adversarial attacks injected by the transmitter, we consider general adversarial mitigation strategies that have been proposed to defend wireless communications classifiers from adversarial attacks as baseline approaches to demonstrate the effectiveness of our proposed method. The selected baselines were chosen as they represent the most widely adopted defense and mitigation strategies in the adversarial machine learning literature, and have been previously applied in analogous RF fingerprinting and wireless security settings \cite{DAE2, MIMODAE}. Each baseline is included to isolate distinct defense principles as outlined below. Specifically, we compare our proposed method to the following baselines: adversarial training \cite{advtrn}, Gaussian smoothing \cite{channeleffectpert1}, frequency domain transferability \cite{dftadvtraining}, amplitude-phase transferability \cite{paadvtraining}, image-based histogram features \cite{past_ai}, and masked autoencoding \cite{masked_auto}. Adversarial training consists of augmenting the training set with gradient-based adversarial samples between batches of training, where the magnitude of the perturbation is empirically determined such that the classifier results in the highest testing accuracy. Here, we consider two adversarial training baselines: one trained using FGSM perturbations generated according to (\ref{l2_fgsm}) and (\ref{linf_fgsm}) and one trained using PGD perturbations generated according to (\ref{l2_pgd}) and (\ref{linf_pgd}). The Gaussian smoothing approach is trained similarly; instead of gradient-based perturbations, however, samples are added with AWGN, where, the mean and variance of the augmented samples are chosen to maximize the testing accuracy. The frequency-based (FQ) model uses the Discrete Fourier Transform (DFT) of its received signals for training. Similarly, the amplitude-phase (PA) model trains on the amplitude and phase features of each received signal. In \cite{past_ai}, received RF signatures are converted into a bivariate histogram image of user-specified dimensions, and each tile is normalized to $[0, 255]$, where pixels with higher values represent a high density of IQ samples and lower values are assigned to tiles with fewer samples. In contrast to using histograms of $224 \times 224$ tiles on captures of 10,000 IQ samples as done in \cite{past_ai}, to avoid training issues caused by very sparse inputs, our significantly less abundant captures of 256 IQ samples prompts us to select a histogram dimension of $100 \times 100$ tiles, which is found to deliver the best performance on this baseline after numerous trials on varying sizes. Masked autoencoding incorporates a mask consisting of a random distribution of 0s and 1s with a parameter that determines the proportion of 0s and 1s \cite{masked_auto}. Prior to training the autoencoder, the RF fingerprint signal is element-wise multiplied with the mask and used as the input sample with the original unmasked counterpart of the signal being the target output. The loss of the autoencoder is calculated between the autoencoder output and the complement of the masked input (in order to identify the error of the 0-element IQ samples) and then minimized.

For all baselines employing augmented training samples, the perturbation magnitude used during training was selected by evaluating multiple PSR levels and choosing the value that maximized classification accuracy while maintaining relatively robust performance across the full PSR range. Multi-$\boldsymbol{\delta}$ training was also evaluated and found to offer marginal improvement at high PSR values at the cost of degraded performance under weak perturbations; the single-$\boldsymbol{\delta}$ selection thus provides a fairer baseline comparison by ensuring all methods achieve comparable unperturbed classification accuracy.

In Fig. \ref{fgsm_res_im}, we show the performance of our approach, in comparison to each considered baseline, in Environment A and Environment B under both $l_{2}$- and $l_{\infty}$-bounded perturbations. From Fig. \ref{fgsm_res_im}, we see that, when the eavesdropper is completely unaware of the perturbation, the transmitter is able to successfully craft a transferable attack to deter the eavesdropper, demonstrated by the eavesdropper's accuracy dropping to as low as 30\%. Our proposed eavesdropper architecture, however, is able to mitigate the effects of the transferable attack, resulting in device identification classification at nearly double the rate (of no resilience) in Environment A and nearly at the same performance as the receiver in Environment B. In contrast, the FGSM and PGD adversarially trained models attain lower RFF classification performance, particularly on high-bounded perturbations, which is the magnitude that the transmitter is likely to use when embedding the perturbation in their transmitted signal as it induces the most potent transferable attack to the eavesdropper (when the eavesdropper is unaware of the perturbation) while delivering strong performance at the receiver.

Moreover, in Environment A and B, our proposed method outperforms each considered baseline. For example, in comparison to adversarial training, our proposed method's performance improves by as high as $40\%$ under the presence of FGSM attacks in both environments. While the transmitted perturbations hinder the unaware eavesdropper at $\text{PSR} <-35$ dB and significantly drops in accuracy at $-20$ dB, our proposed method does not experience impairment until $-20$ dB, with a rebound after a slight dip at $-16$ dB, converging towards $80\%$ in Environment B. The robustness of the FGSM and PGD adversarially trained model starts to degrade after $\text{PSR} = -27$ dB), whereas our method remains unaffected throughout nearly the entire PSR range. Under both $l_{2}$ and $l_{\infty}$ attacks in both environments, our proposed STFT-based classifier at the eavesdropper results in greater robustness than applying Gaussian smoothing for $\text{PSR} \ge -20$ dB. The PA feature representation works well, but has lower overall robustness. Furthermore, when the eavesdropper considers both time and frequency features as in our proposed method, there is a noticeable difference, with respect to the FQ based baseline, in accuracy of as large as $45\%$ and $40\%$ in Environment A and B, respectively, demonstrating that the use of window filters localizes the adversarial perturbations, preventing them from heavily impairing the eavesdropper's robustness. Under the presence of $l_{2}$ FGSM perturbations, Fig. 5 shows the image-based classifier to perform at a consistent accuracy for PSR $< -12$ dB before dropping by as much as $10\%$. In comparison, although our proposed method’s performance declines at weaker perturbation magnitudes, the maximum drop in performance is less than 5 $\%$. When the transmitter injects $l_{\infty}$-bounded FGSM perturbations, the image-based classifier drops $17\%$ in accuracy whereas our proposed STFT approach only degrades in accuracy by $4\%$. Moreover, for $l_{2}$-bounded FGSM perturbations, the masked autoencoder starts to decline in performance near $-17$ dB, whereas our proposed approach declines earlier at $-22$ dB. However, the masked autoencoder continues to decline with increasing perturbation magnitude and drops by more than $10\%$ before settling at $67\%$, while our STFT classifier drops by less than $5\%$ and stops degrading in performance at an earlier PSR in comparison to the masked autoencoder. A similar trend is shown for $l_{\infty}$-bounded FGSM perturbations, where the masked autoencoder decline is more pronounced. Overall, we see from Fig. \ref{fgsm_res_im} that our method outperforms each considered baseline under highly potent attacks $\text{PSR} \geq -22$ dB, indicating an eavesdropper's ability to fingerprint devices transmitting signals injected with FGSM perturbations.   


As illustrated in Fig. \ref{fgsm_res_im}, and since solutions to (\ref{adv_opt:all-lines}), are approximated, (\ref{l2_fgsm}) and (\ref{linf_fgsm}) do not always satisfy (\ref{adv_opt:all-lines}) and induce misclassification. For example, on a single-step attack trained classifier under the presence of FGSM perturbation, instead of proceeding to decline as the magnitude of perturbation rises, the robustness starts to rise. This phenomenon, Catastrophic Overfitting (CO) introduced in \cite{CO2}, suggests that the increased robustness on very strong perturbations is caused by the distortion of the decision boundary. During the adversarial training stage, the classifier may severely overfit to certain examples inducing an over exaggeration in the curvature of the loss function surface \cite{CO1, CO3}. Although the overfitting effect of an FGSM trained eavesdropper achieves higher accuracy than the proposed method and receiver at $\text{PSR} \ge -10$ dB, very high PSRs are unrealistic due to the power budget constraint at the transmitter. In addition, the ``turning point" at which overfitting starts to occur is arbitrary, leading to uncertain eavesdropping performance in a desired interval of PSRs.

\begin{figure}[t] 
	\centering
	\includegraphics[width=\halfwidth\columnwidth]{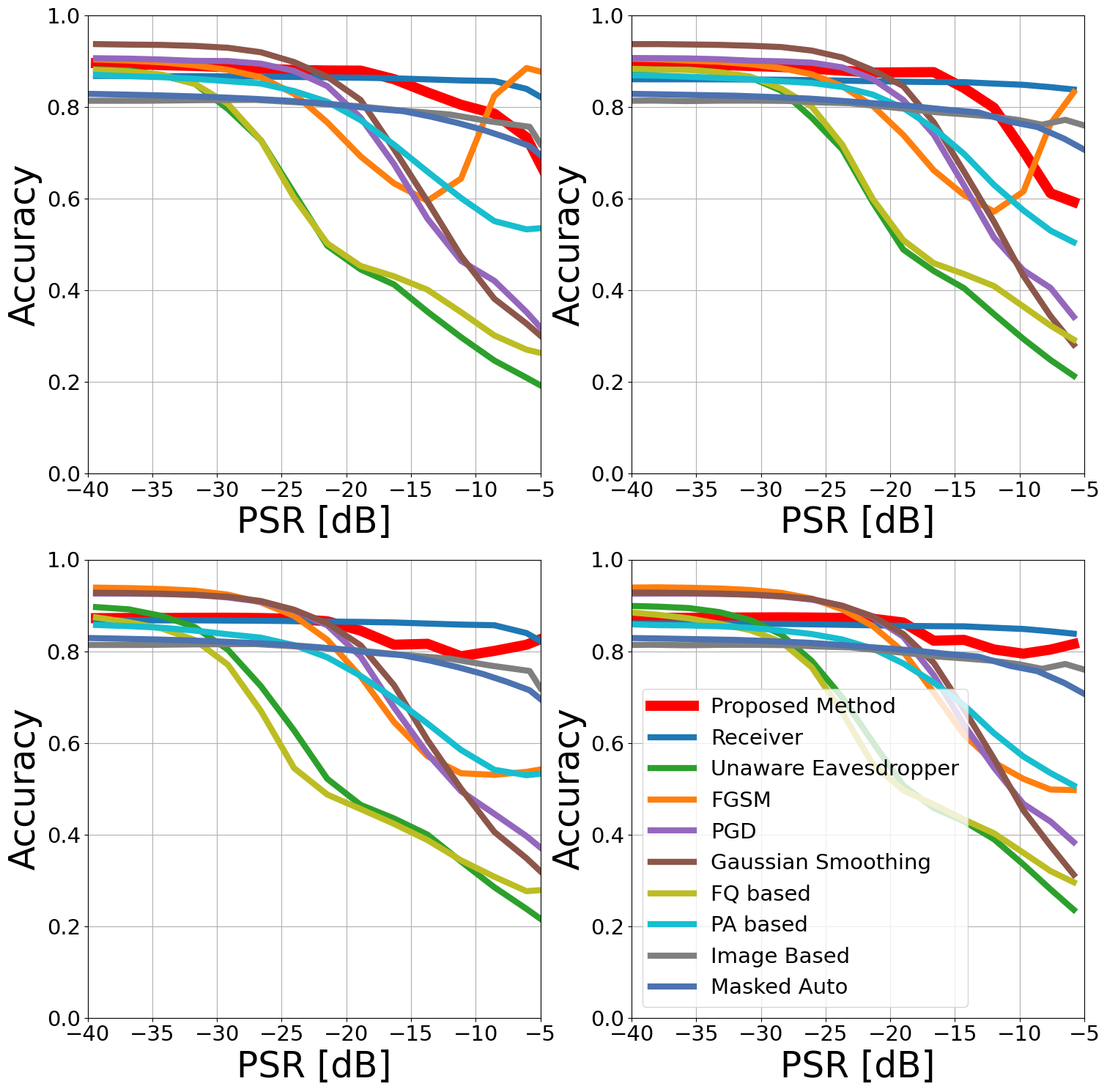}
	\caption{Eavesdropper's classification performance against PGD perturbations injected by the transmitter in comparison to the receiver, unaware eavesdropper, and each considered baseline under $l_{2}$- and $l_{\infty}$-bounded perturbations in Environment A (top left and top right), and $l_{2}$- and $l_{\infty}$-bounded perturbation in Environment B (bottom left and bottom right).  Similar to Fig. \ref{fgsm_res_im}, under our proposed method, the eavesdropper outperforms each considered baseline and again approaches the performance of the receiver in certain instances.}
	\label{pgd_res_im}
\end{figure}

\subsection{Eavesdropper Resilience Against Iterative Perturbations} \label{pgd_res}
We now evaluate our proposed method under PGD adversarial attacks injected by the transmitter. As shown in Fig. \ref{pgd_res_im}, when the eavesdropper is unaware of the presence of an attack injected by the transmitter, in both Environment A and Environment B, the accuracy decline begins as early as $\text{PSR} = -33$ dB and steadily decreases to $20\%$. Our proposed method, on the other hand, in Environment A does not decline in performance until a much stronger perturbation of nearly $-19$ dB PSR, falling to an accuracy of 60\%, providing a maximum improvement in accuracy of $45\%$ at $-15$ dB. In Environment B, our proposed method is able to withstand the effect of the adversarial attack injected by the transmitter across nearly the entire considered PSR range. In contrast to the unaware eavesdropper's steady decline in robustness, our implementation endures less dramatic slopes (the unaware eavesdropper experiences a $20\%$ drop in accuracy compared to our method only suffering by $5\%$) and convergences at $80\%$.

In Environment A and B, at moderate PSRs ($-23$ dB), our method maintains an increasing disparity over the FGSM trained eavesdropper in the $l_{2}$ and $l_{\infty}$ cases. A similar phenomenon appears in a multi-step attack trained classifier where the slope of the PGD trained model is steep, experiencing a significant decline in robustness, in contrast to our proposed method. With respect to the performance of the Gaussian smoothed model, although the baseline performance in the absence of an adversarial perturbation is slightly higher than our method, after a moderate PSR ($-21$ dB), our method achieves greater robustness, which can be observed by the widening gap, where Environment B has a more apparent improvement. In comparison with the existing baselines, our proposed method achieves significantly higher robustness than the FQ based (time-agnostic) model (as high as 45\%  and 60\% in Environment A and B, respectively) and the PA based model (as high as 22\%  and 31\% in Environment A and B, respectively). Similar improvements over the image-based RFF baseline are observed by our proposed framework and are consistent with the improvements experienced for FGSM perturbations, albeit to a slightly lesser extent as expected due to the higher potency of the PGD perturbation. The masked autoencoder also demonstrates consistent performance in comparison to signals injected with FGSM perturbations. When considering the receiver's ability to recover the perturbed transmission, our proposed method allows the eavesdropper to achieve an even greater robustness than the receiver for weak PSRs ($\text{PSR} \leq-20$ dB) in both environments. Comparing Fig. \ref{pgd_res_im} with Fig. \ref{fgsm_res_im}, the robustness of our proposed method remains consistent against a perturbed transmission regardless of its signal environment and type of perturbation.

\begin{figure}[t] 
	\centering
	\includegraphics[width=0.9\columnwidth]{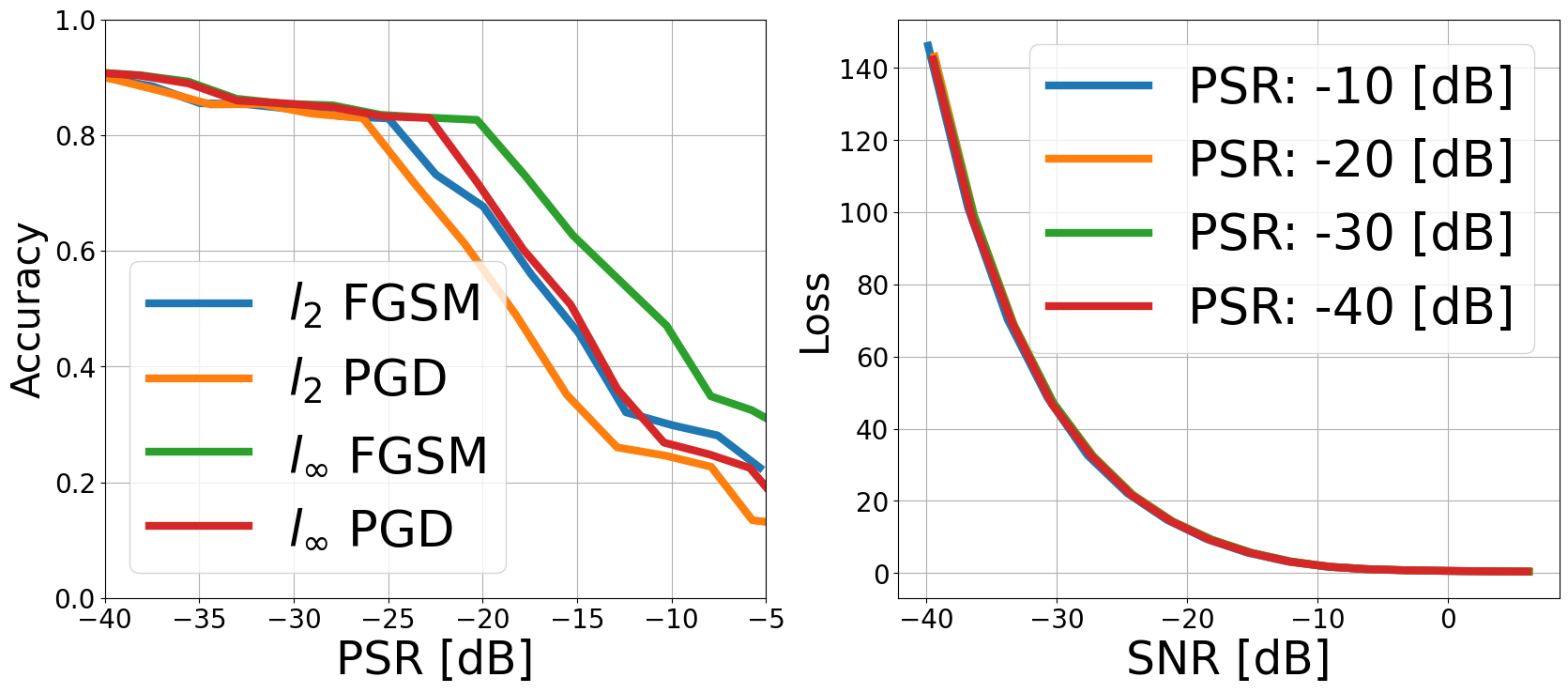}
	\caption{Left: Eavesdropper's classification performance on perturbed signals crafted at the transmitter with full knowledge of the eavesdropper's classification framework. Here, we see that the transmitter is able to effectively induce misclassification at the eavesdropper with increasing potency when the transmitter is aware of the eavesdropper's RFF framework. Right: Receiver's loss values on signals with adversarial interference and varying signal-to-noise ratios (SNR). Here, as expected, we see that as the SNR increases, classification performance increases.} 
	\label{dae_whitebox_attack_img}
\end{figure}

\subsection{Ablation Studies}\label{whitebox_attack}

In this section, we independently assess the performance of the eavesdropper and receiver under varying assumptions. We first consider an eavesdropper-aware transmitter and then we consider a signal-to-noise ratio (SNR)-varying channel between the transmitter and receiver.

We first consider the performance of our proposed method when the transmitter has full information of the eavesdropper's RFF scheme and architecture and crafts an attack accordingly. In this capacity, the transmitter generates adversarial attacks with full knowledge about the presence of the eavesdropper and its classifier, including its STFT scheme. Specifically, at the transmitter, an adversarial attack is generated on the STFT features themselves using the gradient of the classifier deployed at the eavesdropper. After obtaining the perturbed features in the STFT domain, the transmitter calculates the inverse STFT in order to model the perturbation in the time-domain. The resulting signal is then transmitted. Since the transmitter is aware that the eavesdropper will fingerprint the intercepted signal using the STFT, this baseline attack is more likely to induce misclassification at the eavesdropper. We perform this experiment for each considered attack across multiple potencies and show our results in Fig. \ref{dae_whitebox_attack_img} (left). As shown in Fig. \ref{dae_whitebox_attack_img} (left), and as expected by the transmitter, the accuracy of the eavesdropper degrades to low accuracy as the potency of the perturbation increases. This result reinforces that the transmitter is able to fool an eavesdropper when the transmitter is aware of how the eavesdropper operates. However, under our method, in the case where the transmitter is unaware that the eavesdropper is leveraging both time and frequency components for classification, the eavesdropper can evade misclassification, as we justify in Theorem \ref{thm1} and show empirically in Sec. \ref{fgsm_res} -- \ref{pgd_res}. In addition, we note that the DAE is trained at a single PSR of $-20$~dB, selected empirically as the value that best balances reconstruction quality across the full PSR range after evaluating multiple training levels. The DAE's generalization to other PSR  levels is evidenced by the consistently high receiver classification performance shown across the full PSR range in Figs.~\ref{fgsm_res_im} and~\ref{pgd_res_im}, since the receiver uses the DAE to filter additive perturbations throughout all evaluations.

We next consider an SNR-varying channel between the transmitter and receiver in order to assess the performance of the receiver under different noise conditions in the wireless channel. Specifically, we evaluate the receiver's performance with varying SNR values at multiple PSRs on each considered perturbation design. Our results for $l_\infty$-bounded PGD perturbations are shown in Fig. \ref{dae_whitebox_attack_img} (right). Each of the other considered perturbations exhibited a similar trend, so we omit them here for brevity. From Fig. \ref{dae_whitebox_attack_img} (right), we see that as the SNR decreases, the loss of the receiver's classifier increases. This trend is consistent with the performance of deep learning classifiers on wireless systems, where such models fail to attain high performance on low SNR signals. However, as SNR increases, the performance of deep learning models also increases as shown in Fig. \ref{dae_whitebox_attack_img} (right). Our framework's consistent performance with general deep learning-based signal processing reinforces its correct operation at the receiver.

\section{Conclusion} \label{conclusion_sec}

In an effort to mislead eavesdroppers from performing deep learning-based device authentication, while ensuring receivers are not misled, transmitters may inject adversarial perturbations into their broadcast signals. In this work, we developed a robust eavesdropping framework in which the eavesdropper jointly uses time and frequency components to mitigate the effects of such transferable adversarial attacks.  We theoretically showed that, on the spatial-temporal features of the perturbed transmission, the effects of the perturbation are localized in regions of the spectrogram that have negligible influence on the classifier. Using a real-world device authentication dataset, we empirically validated our approach and showed its robust performance in comparison to several baseline approaches. Overall, our work reveals that transferable adversarial perturbations are not sufficient for deterring eavesdroppers in deep learning-based wireless communications when the transmitter is unaware of the eavesdropper's time--frequency defense. When the transmitter has full knowledge of the eavesdropper's STFT-based architecture, white-box attacks crafted directly on the STFT features can effectively degrade eavesdropper performance, as demonstrated in Section~\ref{whitebox_attack}. This distinction provides a clear delineation of the operating regime of our proposed framework.

Our proposed framework can also be generalized and adopted across different wireless systems. For example, in Automatic Modulation Classification (AMC), intercepted signals can be analyzed to identify the modulation scheme, enabling adaptive demodulation or threat assessment in cognitive and security applications. In Direction of Arrival (DoA) estimation, spatial information from the intercepted waveforms, often captured using antenna arrays, can be processed to locate the transmitter’s position, supporting tasks such as source tracking or beamforming. For Multiple Input Multiple Output (MIMO) systems, eavesdropping can reveal channel state information (CSI) across multiple spatial paths, allowing for characterization of multipath propagation, antenna correlations, and spatial diversity patterns. Together, these capabilities demonstrate how passive interception can feed into higher-level wireless intelligence, from classifying signal types to locating emitters and exploiting spatial channels for advanced communications or countermeasure strategies. In future work, we anticipate exploring how susceptible such deep learning-based wireless systems are to circumventing attempts made by a transmitter to avoid eavesdropping in different wireless networks.

\textbf{Ethical Considerations:} This work reveals that passive eavesdroppers equipped with a time--frequency-based RFF classifier can effectively fingerprint transmitters even when the transmitter injects adversarial perturbations designed to prevent recognition. While this result has legitimate applications in network security auditing and resilience testing, it also has potential dual-use implications. We emphasize that this paper is intended to surface a previously underappreciated vulnerability so that the community can develop stronger countermeasures. In this spirit, we suggest two complementary mitigation directions for transmitters seeking stronger anti-eavesdropping guarantees: (i) instead of relying on black-box transferability, transmitters should craft adversarial perturbations with explicit knowledge of the eavesdropper's feature representation (e.g., targeting the STFT domain directly, as demonstrated in Section~\ref{whitebox_attack}); and (ii) physical-layer security mechanisms such as directional beamforming, cooperative jamming, or frequency-hopping can be combined with adversarial perturbations to reduce the eavesdropper's received signal quality and thus further limit its classification capability.


\bibliography{references}

\bibliographystyle{IEEEtran}

\section*{Appendix}

\subsection{Proof of Theorem \ref{thm1}} \label{app_proof}

\begin{proof}
 The received signal at the eavesdropper can be written as $s_E = y + \psi + g$, where $y = \mathbf{H}_E\mathbf{x}$, $\psi = \mathbf{H}_E\pmb{\delta}$, and $g$ is AWGN. By linearity of the STFT,
\begin{equation}
S_E^{(n,f)} = Y^{(n,f)} + \Psi^{(n,f)} + G^{(n,f)}, \label{stft_linear}
\end{equation}
where $Y=\mathcal{S}[y]$, $\Psi=\mathcal{S}[\psi]$, and $G^{(n,f)}=\mathcal{S}[g]$. Let $S_E^{\prime(n,f)} = Y^{(n,f)} + G^{(n,f)}$ denote the STFT of the received signal when $\pmb{\delta} = \mathbf{0}$ (i.e., $\Psi^{(n,f)} = 0$). Expanding the PSD using \eqref{stft_linear} and subtracting $|S_E^{\prime(n,f)}|^2$ yields the exact identity
\begin{equation}
|S_E^{(n,f)}|^2 - |S_E^{\prime(n,f)}|^2 = 2\,\mathrm{Re}\big[S_E^{\prime(n,f)}\,
\overline{\Psi^{(n,f)}}\big] + |\Psi^{(n,f)}|^2. \label{exact_diff}
\end{equation}
Applying the triangle inequality, $|\mathrm{Re}[z]| \leq |z|$, and $|S_E^{\prime(n,f)}| \leq |Y^{(n,f)}| + |G^{(n,f)}|$ to \eqref{exact_diff} gives
\begin{equation}
\Big|\,|S_E^{(n,f)}|^2 - |S_E^{\prime(n,f)}|^2\,\Big| \leq \nonumber 
\end{equation}
\begin{equation}
2\big(|Y^{(n,f)}|+|G^{(n,f)}|\big)|\Psi^{(n,f)}| + |\Psi^{(n,f)}|^2. \label{step1_bound}
\end{equation}

Now, we bound $|\Psi(\hat{f})|$ via the power budget). By Parseval's theorem, the energy of $\psi[k] = (\mathbf{H}_E\pmb{\delta})[k]$ satisfies $\sum_k |\psi[k]|^2 \leq \|\mathbf{H}_E\|_2^2\|\pmb{\delta}\|_2^2 \leq \|\mathbf{H}_E\|_2^2 P_T$, where the final inequality follows from the power-budget constraint $\|\pmb{\delta}\|_2^2 \leq P_T$ in (\ref{adv_opt:line_4}). Since the discrete-time Fourier transform satisfies $|\Psi(\hat f)|^2 \leq \sum_k|\psi[k]|^2$ for all $\hat f$, it follows that
\begin{equation}
|\Psi(\hat f)| \leq \sqrt{P_T}\,\|\mathbf{H}_E\|_2 \quad \text{for every } \hat f.
\label{psi_bound}
\end{equation}

Focusing on the spectral leakage of $\Psi$ into the mid-band), the STFT of $\psi$ can be written as
\begin{equation}
\Psi^{(n,f)} = \sum_{\hat f=-\infty}^{\infty} \Psi(\hat f)\,W(\hat f - f)\,
e^{-j2\pi(\hat f - f)n}. \label{psi_stft_full}
\end{equation}
By Assumption~1 (or Assumption~2), $\Psi(\hat f) \approx 0$ outside the band $(f_\psi, f_\psi+\Delta_\psi)$, so \eqref{psi_stft_full} reduces to a sum over $\hat f \in (f_\psi, f_\psi+\Delta_\psi)$. Since the Gaussian window's spectrum satisfies $|W(\hat f - f)| \leq Ke^{-(\hat f - f)^2/(2\sigma^2)}$ for some constant $K>0$, and since $|\hat f - f| \geq \kappa\sigma$ for every $\hat f$ in the perturbation band and every $f \in (f_a,f_b)$,
\begin{equation}
|W(\hat f - f)| \leq K e^{-\kappa^2/2} \quad \text{for all such } \hat f, f.
\label{boundingterm}
\end{equation}
Combining \eqref{psi_bound} and \eqref{boundingterm} and summing \eqref{psi_stft_full} over the (finite-width) perturbation band,
\begin{equation}
|\Psi^{(n,f)}| \leq \sqrt{P_T}\,\|\mathbf{H}_E\|_2 \cdot K e^{-\kappa^2/2} \cdot \Delta_{\psi}
\quad \text{for } f \in (f_a, f_b). \label{psi_nf_bound}
\end{equation}

Finally, combining the bounds and substituting \eqref{psi_nf_bound} into \eqref{step1_bound} gives, for $f \in (f_a,f_b)$,
\begin{equation}
\Big|\,|S_E^{(n,f)}|^2 - |S_E^{\prime(n,f)}|^2\,\Big| \leq\  K^2 P_T \|\mathbf{H}_E\|_2^2 \Delta_{\psi}^{2} e^{-\kappa^2} +\nonumber
\end{equation}
\begin{equation}
2K\sqrt{P_T}\,\|\mathbf{H}_E\|_2\, e^{-\kappa^2/2}\Big(|Y^{(n,f)}| + |G^{(n,f)}|\Big),
\end{equation}
which is exactly \eqref{thm1_bound}. Since $K$, $P_T$, $\|\mathbf{H}_E\|_2$, $|Y^{(n,f)}|$, and $|G^{(n,f)}|$ are all finite, the right-hand side is a finite, well-defined upper bound for every finite $\kappa$, and clearly tends to zero as $\kappa \to \infty$ since both $e^{-\kappa^2/2} \to 0$ and $e^{-\kappa^2} \to 0$. An identical argument applies under Assumption~2, with $\kappa$ replaced by $\mu$ and the perturbation band located near $f_a$ rather than $f_b$, yielding the same bound with $\mu$ in place of $\kappa$.
\end{proof}

\noindent \textbf{Andrew Yuan} is currently pursuing his B.S. in Electrical Engineering at the University of California San Diego. His research interests lie in wireless communications, adversarial deep learning, and controls. \\

\noindent \textbf{Rajeev Sahay} received the B.S. degree in electrical engineering from The University of Utah, Salt Lake City, UT, USA, in 2018, and the M.S. and Ph.D. degrees in electrical and computer engineering from Purdue University, West Lafayette, IN, USA, in 2021 and 2022, respectively. Currently, he is a faculty member in the Department of Electrical and Computer Engineering at the University of California, San Diego (UCSD). He was the recipient of the Purdue Engineering Dean’s Teaching Fellowship, the UCSD best undergraduate teaching award, and was named an Exemplary Reviewer by the IEEE Wireless Communications Letters. His research interests lie in the intersection of networking and machine learning, especially in their applications to wireless communications and engineering education. 

\end{document}